\documentclass[a4paper,11pt]{amsart}

\usepackage{graphicx}
\usepackage{a4wide}
\usepackage[dvipsnames]{xcolor}
\usepackage{epstopdf, epsfig}
\usepackage{amsaddr}
\usepackage{amsmath,amsfonts,amssymb,amsgen,amsbsy,dsfont}
\usepackage{graphicx}
\usepackage{subfig}
\usepackage{upgreek}
\usepackage{bm}
\usepackage{hyperref}
\usepackage{interval}
\usepackage{cleveref}
\usepackage{calrsfs}
\usepackage{yhmath}
\usepackage[round]{natbib}

\DeclareMathAlphabet{\pazocal}{OMS}{zplm}{m}{n}

\newcommand{\rf}[1]{(\ref{#1})}

\newcommand\Rey{\pazocal{R}}  
\newcommand\Bos{\pazocal{B}_{\textrm{s}}}  
\newcommand\Bob{\pazocal{B}_{\textrm{b}}}  
\newcommand\Fr{\pazocal{F}}  

\DeclareMathOperator{\sech}{sech}

\newcommand{\ud}{\mathrm{d}}

\newcommand{\ue}{\mathrm{e}}
\newcommand{\ui}{\mathrm{i}}

\renewcommand{\Re}{\operatorname{Re}}

\begin{document}

\title[The instability of a membrane enclosed by two viscous fluids]{The instability of a membrane enclosed by two viscous fluids with a free surface}

\author[J. Labarbe]{Joris Labarbe}
\address{Universit\'e C\^ote d'Azur, CNRS UMR 7351,  Laboratoire J.~A. Dieudonn\'e, 
Parc Valrose, 06108 Nice cedex 2, France}
\email{joris.LABARBE@univ-cotedazur.fr}


\begin{abstract}
This study examines the stability of a flexible material interface between two fluids of the same viscosity in interaction with a free surface.
When the layers are motionless, we provide evidence for the onset of a novel instability by means of analytical and numerical solution of the associated boundary value problem in the region stable against Rayleigh--Taylor instability, i.e. when the acceleration due to gravity acts from the lighter to the heavier fluid.
This destabilisation phenomenon is attributed to the non-conservative tangential forces acting at the interface and the fluid-structure interaction.
Furthermore, we examine the scenario in which an external forcing mechanism induces a monotonic parallel shear flow within the upper layer.
In addition to the long-established inflectional instability predicted in the inviscid limit, we demonstrate the existence of membrane flutter in the absence of density stratification.
The latter is either due to an over-reflection process of surface gravity waves or to the growth of Tollmien--Schlichting waves, as outlined in the context of boundary-layer theory.
This fluid-structure configuration represents a paradigmatic model for investigating the interplay between inflectional, radiation-induced and shear-induced instabilities. 
It also serves as a viscous counterpart to the classical Kelvin--Helmholtz instability when layers with distinct densities are assumed.
\end{abstract}

\maketitle

\section{Introduction} \label{sec0}

The interaction between a moving fluid and a solid represents one of the most significant research topics in classical physics since the advent of Newtonian mechanics.
It is of great importance to investigate this coupled dynamics, as it is responsible for the majority of fundamental real-world phenomena observed on a macroscopic scale.
Fluid-structure interaction enables the development of reliable and efficient designs in a range of applied fields, including aerospace engineering (aircraft wings immersed in supersonic flows), civil engineering (buildings or bridges exposed to wind and seismic events), biomedical engineering (blood flows streaming through cardiovascular devices), and environmental engineering (wind turbines producing renewable energy).
In most cases, an essential part of the research on this subject is devoted to the theoretical modelling of the solid body when coupled to a moving fluid \citep{DH01}.

In a celebrated experiment on the incident flow past a cylinder, \cite{VK12} discovered the existence of vortex streets in the wake of the solid.
This phenomenon is attributed to the detachment of the viscous boundary layer when the inertial effects overcome the friction at the circular wall or equivalently, when the Reynolds number exceeds a critical value \citep{S79}.
Since then, the boundary-layer stability has attracted the attention of numerous scientists and was notably applied to investigate the transition to turbulence in wall-bounded shear flows.
It was Prandtl who first demonstrated that small wave disturbances, when interacting with a leading edge, undergo an unstable mechanism due to the friction layers near the wall \citep{Lamb}.
Under linear theory, i.e. when perturbations are assumed as infinitesimal, the analysis of the Orr--Sommerfeld equation \citep{O07} predicts the growth of Tollmien--Schlichting waves induced by viscosity until they ultimately collapse into vortices \citep{B96}.
This phenomenon still exists at a flexible boundary, although the compliant response of the structure is usually acting as a stabilisation effect \citep{B60}.

The study of interface fluctuations of multiple-phase fluids or binary mixtures represents an ongoing and significant research area, notably in mixing experiments involving active materials \citep{Davies}.
A substantial portion of this research is dedicated to compute the stability of interfacial waves when excited by mechanical forces or by the interaction with a surrounding fluid and the role they play in the dynamics of the system.
For example, high-resolution spectroscopy is employed within light scattering experiments to measure the viscoelastic response of thin interfaces \citep{K71}.
In the field of cellular fluid mechanics, thin membranes are considered as an expository model to examine the interactions between lipids and proteins at the interface between two fluids \citep{K02}.
It is also of significant interest in the critical context of global warming to consider the topic of ocean waves and sea ice interactions, as this can help to predict the dynamics of ice floes in polar regions \citep{S20}.
It should be noted that the aforementioned applications do not represent an exhaustive list of the fluid-structure interaction problems. 
Rather, they are intended to illustrate a selection of stimulating ongoing research topics.

A paradigmatic wave-system to exhibit the phenomenon of radiation damping experienced by some vibrating bodies is the so-called Lamb oscillator, named after its instigator \citep{L00}.
This one-dimensional model describes the motion of a spring-mass system that is coupled to an infinite string, which is attached at the point mass.
When departing from its equilibrium state, the oscillator is subject to effective dissipation induced by the travelling waves propagating outwards, resulting in a decay of its amplitude over time.
Variants of this system have been studied (notably by adding gyroscopic forces) to introduce the concept known as radiation-induced instability \citep{BKMR94}.
This phenomenon arises in some oscillatory systems where the emission of waves eventually builds up energy by destroying the gyroscopic stabilisation, in an analogous manner to the addition of small dissipation to canonical gyroscopic systems \citep{KM07}.
Meanwhile, this instability mechanism has been employed to explain classical hydrodynamic instabilities, such as the Kelvin--Helmholtz mechanism, through the concept of negative energy waves \citep{C79}.

An interesting problem related to the fluid-structure interaction of an inviscid surface layer with a material interface is the so-called Nemtsov problem \citep{N85}.
Originally, Nemtsov suggested this fundamental model to relate the flutter instability of a flexible membrane to the radiation of surface gravity waves (of negative energy) within the region of anomalous Doppler effect \citep{N76}.
At that time, it was well-established that radiation by uniformly moving sources could yield some surprising phenomena in particle physics via the Vavilov--Cherenkov effect \citep{G96}.
Historically, the Nemtsov problem was the first application of negative energy waves, radiation-induced instability and anomalous Doppler effect in the field of fluid mechanics.
Recently, this problem was revisited by \cite{LK20} and \cite{LK22}, who have taken into account the finite size effects of the fluid layer and the membrane chord length, respectively.
However, in both studies, the flow was considered to be uniform, irrotational and inviscid.
Allowing for the viscous effects to be present and assuming a shear motion, the stability of the system can largely be altered, as evidenced by the Faraday problem \citep{KT94} or the viscous counterpart of the Kelvin--Helmholtz instability.

This article presents a generalisation of the Nemtsov problem by investigating the stability of a thin membrane enclosed by two viscous fluids with a free surface.
Section \ref{sec1} introduces the governing equations, the associated boundary conditions, the choice of the equilibrium state, as well as the analytical and numerical methods used throughout the paper.
We present in section \ref{sec2} the stability of the motionless configuration, i.e. when the background flow is absent.
We demonstrate here the existence of a new instability by means of the coupled dispersion relation and the numerical solution of the Navier--Stokes equations.
In section \ref{sec3} we examine the influence of a monotonic shear flow generated by an external forcing within the upper layer.
We describe the various unstable mechanisms, notably the modes of inflectional-induced, radiation-induced and viscosity-induced instability.
Finally, we discuss on the results and the future extensions of this work in section \ref{sec4}.

\section{Formulation of the problem} \label{sec1}

\begin{figure}[t!]
    \centering
    \includegraphics[width=.8\textwidth]{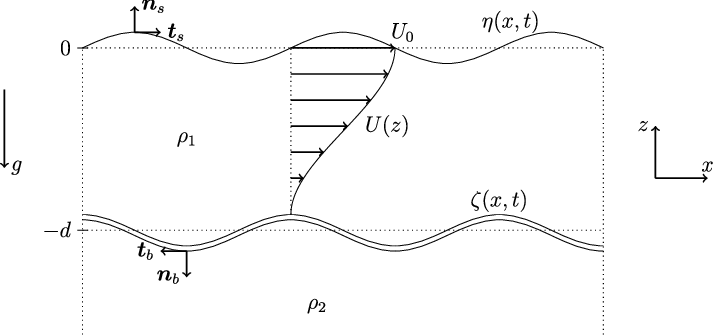}
    \caption{Schematic representation of a viscous shear layer with an infinite horizontal extension enclosed by a free surface above and a material interface underneath (the system is not represented at rest intentionally). 
    Below the membrane lies a motionless fluid of the same viscosity as the upper layer but not necessarily the same density. 
    The thickness of the membrane is illustrated for visual interests although we neglect its influence.}
    \label{Fig1}
\end{figure}

We consider a flexible material interface (e.g. a membrane) with density per unit area $\rho_m$ and  negligible thickness being located between a free surface layer of constant mean depth $d$ and a semi-infinite fluid layer (this system is depicted in figure \ref{Fig1}).
The membrane is rigidly fixed at its extremities (prohibiting tangential motion of the interface) although the chord length is significantly larger than the layer height, such that the horizontal extension is supposed infinite.
We assume the two fluids to have the same kinematic viscosity $\nu$ (we could easily relax this statement) but distinct densities $\rho_1$ and $\rho_2$ associated with the upper and lower layer, respectively.
Above the surface lies an inviscid fluid moving uniformly at velocity $U_0$ and constant pressure $P_0$ serving as a model for, e.g., wind-induced drift currents.
Finally, gravity acts downwards and is designated by the uniform acceleration rate $g$.
We notice that this configuration is reminiscent of Rayleigh--Taylor and Kelvin--Helmholtz problems when both layers are at rest or if the upper layer is in parallel motion, respectively \citep{Chandra}.

\subsection{Governing equations and boundary conditions}

We express lengths, time, velocity, density and relative pressure (with respect to the value of reference $P_0$) in the units of $d$, $\sqrt{d/g}$, $\sqrt{gd}$, $\rho_2$ and $\rho_2 g d$, respectively.
As a consequence, all expressions introduced hereafter are given in their non-dimensional forms.

This hydrodynamical system is separated into two distinct domains described in Cartesian coordinates, denoted as $\Omega_1=\{(x,z):x\in\mathbb{R},-1+\zeta(x,t) \leq z \leq \eta(x,t)\}$ for the upper layer and $\Omega_2=\{(x,z):x\in\mathbb{R},-\infty \leq z \leq -1+\zeta(x,t)\}$ for the lower layer.
We introduced in the latter the surface elevation $\eta(x,t)$ as well as the membrane displacement $\zeta(x,t)$ while assuming both interfaces as flat when at rest (this particular choice dictates the equilibrium state introduced in a subsequent section).
From the balance of linear momentum and the mass conservation \citep{Lamb}, the velocity field $\bm{u}_j=(u_j,w_j)$ and pressure $p_j$ (defined in their respective domains $\Omega_j$) are governed by the incompressible Navier--Stokes equations
\begin{align}
\label{eom1}
\partial_t \bm{u}_j + (\bm{u}_j \bm{\cdot} \bm{\nabla}) \bm{u}_j &= - \bm{\nabla} p_j - \alpha_j \bm{e}_z + \bm{f} + \Rey^{-1} \Delta \bm{u}_j , \\
\label{eom2}
\bm{\nabla} \bm{\cdot} \bm{u}_j &= 0 ,
\end{align}
where $\bm{\nabla}$ is the del operator and $\Rey=(\sqrt{gd}d)/\nu$ is the Reynolds number.
The scaled densities of the fluids lying above and below the material interface are given by $\alpha_1\equiv\alpha_f=\rho_1/\rho_2$ and $\alpha_2 = 1$, respectively.
We note that $\alpha_f<1$ describes a configuration stable against the Rayleigh--Taylor instability mechanism \citep{R80,T50}.
The restoring gravitational force in \rf{eom1} is accounted for by the pressure effects and the term $\bm{f}=(f,0)$ describes an external forcing, balancing the viscous forces and sustaining a prescribed steady velocity profile $U(z)$ within the upper layer.
We introduce the base state solution along with further details on its particular choice in the next section.

The boundary conditions associated with the deformable interfaces are the kinematic conditions, supplemented with normal and tangential (due to frictional effects) stress conditions.
The former, expressing the continuity of the velocity field across the interfaces, are expressed in terms of surface displacements as
\begin{alignat}{3}
\label{kc1}
& u_1 = \Fr, \quad && w_1 = \left( \partial_t + \Fr \partial_x \right) \eta, \quad && \textrm{at}\:\: z = \eta(x,t) , \\
\label{kc2}
& u_1 = u_2 = 0, \quad && w_1 = w_2 = \partial_t \zeta, \quad && \textrm{at}\:\: z = -1+\zeta(x,t) ,
\end{alignat}
where the Froude number $\Fr=U_0/\sqrt{gd}$ measures the magnitude of the base flow against the characteristic phase speed of (non-dispersive) surface gravity waves in shallow water.

Regarding the dynamical boundary conditions, we follow the work of \cite{D74} and express separately the normal and tangential projections of the linear momentum jump conditions.
Doing so requires to introduce the Cauchy stress tensor within the two fluid layers \citep{Lamb}, given by
\begin{equation}
\bm{\tau}_j = - p_j \bm{I} + 2 \alpha_j \Rey^{-1} \bm{E}_j = - p_j \bm{I} + \alpha_j \Rey^{-1} \left[ (\bm{\nabla} \otimes \bm{u}_j) + (\bm{\nabla} \otimes \bm{u}_j)^{T} \right],
\end{equation}
where $\bm{I}$ denotes the identity matrix, $\bm{E}$ the strain rate tensor (accounting for the effect of viscosity) and $T$ the transpose operator.
Hereafter, we note the variables evaluated at the surface, i.e. at the location $z=\eta(x,t)$, with the subscript `s' while the variables evaluated at the bottom, i.e. at $z=-1+\zeta(x,t)$, are written with the subscript `b'.
At the free surface where the fluid above is assumed inviscid, motionless and with constant pressure, the jump conditions reduce to
\begin{align}
\label{dbcn1}
p_1 - 2 \alpha_f \Rey^{-1} \bm{n}_{\textrm{s}} \bm{\cdot} ( \bm{E}_1 \bm{\cdot} \bm{n}_{\textrm{s}} ) &= \alpha_f \Bos \left( \bm{\nabla}_{\textrm{s}}^{\parallel} \bm{\cdot} \bm{n}_{\textrm{s}} \right) , \\
\label{dbct1}
\bm{t}_{\textrm{s}} \bm{\cdot} ( \bm{E}_1 \bm{\cdot} \bm{n}_{\textrm{s}} ) &= 0 ,
\end{align}
with $\Bos=\sigma_{\textrm{s}}/(\rho_1gd^2)$ the surface Bond number (we suppose the surface tension $\sigma_{\textrm{s}}$ as constant) and where the tangential del operator at the surface is defined by $\bm{\nabla}_{\textrm{s}}^{\parallel} = (\bm{I} - \bm{n}_{\textrm{s}} \otimes \bm{n}_{\textrm{s}}) \bm{\cdot} \bm{\nabla}$ \citep{Aris}.
In terms of the surface elevation, the outward-oriented normal and tangent unit vectors (cf. figure \ref{Fig1}) are expressed, respectively, as 
\begin{equation}
\bm{n}_{\textrm{s}} = \frac{(-\partial_x\eta,1)^{T}}{\sqrt{1 + (\partial_x\eta)^2}} , 
\quad 
\bm{t}_{\textrm{s}} = \frac{(1,\partial_x\eta)^{T}}{\sqrt{1 + (\partial_x\eta)^2}} .
\end{equation}

We obtain the boundary conditions at the material interface by applying the same procedure, although the jump in linear momentum now involves the inertia and inner weight of the structure.
Assuming the membrane has an intrinsic tension $\sigma_\textrm{b}$ and no tangential velocity (since it is rigidly fixed at its extremities), it yields
\begin{align}
\label{dbcn2}
(p_1-p_2) - 2 \Rey^{-1} \bm{n}_{\textrm{b}} \bm{\cdot} ( \alpha_f \bm{E}_1 \bm{\cdot} \bm{n}_{\textrm{b}} - \bm{E}_2 \bm{\cdot} \bm{n}_{\textrm{b}} ) &= \alpha_m \left( A_m^{\perp} + \bm{n}_{\textrm{b}}\bm{\cdot}\bm{e}_z \right) + \Bob \left( \bm{\nabla}_{\textrm{b}}^{\parallel} \bm{\cdot} \bm{n}_{\textrm{b}} \right) , \\
\label{dbct2}
-2 \Rey^{-1} \bm{t}_{\textrm{b}} \bm{\cdot} ( \alpha_f \bm{E}_1 \bm{\cdot} \bm{n}_{\textrm{b}} - \bm{E}_2 \bm{\cdot} \bm{n}_{\textrm{b}} ) &= \alpha_m \left( A_m^{\parallel} + \bm{t}_{\textrm{b}}\bm{\cdot}\bm{e}_z \right) ,
\end{align}
where $\Bob=\sigma_\textrm{b}/(\rho_2gd^2)$ is the material Bond number, $\alpha_m=\rho_m/(\rho_2d)$ the added mass ratio and $(A_m^{\perp},A_m^{\parallel})$ are the interface acceleration components (given by expressions \rf{Amn} and \rf{Amt} from Appendix \ref{app0}).
The differential operator $\bm{\nabla}_{\textrm{b}}^{\parallel} = (\bm{I} - \bm{n}_{\textrm{b}} \otimes \bm{n}_{\textrm{b}}) \bm{\cdot} \bm{\nabla}$ represents the counterpart of $\bm{\nabla}_{\textrm{s}}^{\parallel}$ at the bottom.
As before, we express the normal and tangent unit vectors at the material interface (oriented downwards from the upper layer, as depicted in figure \ref{Fig1}) in terms of its displacement, that is
\begin{equation}
\label{nbtb}
\bm{n}_{\textrm{b}} = \frac{(\partial_x\zeta,-1)^{T}}{\sqrt{1 + (\partial_x\zeta)^2}} , 
\quad 
\bm{t}_{\textrm{b}} = \frac{(-1,-\partial_x\zeta)^{T}}{\sqrt{1 + (\partial_x\zeta)^2}} .
\end{equation}

The set of equations to solve is thus given by the Navier--Stokes system \rf{eom1}--\rf{eom2} with boundary conditions \rf{kc1}--\rf{kc2}, \rf{dbcn1}--\rf{dbct1} and \rf{dbcn2}--\rf{dbct2}, supplemented with the evanescence condition $\bm{u}_2\to\bm{0}$ at $z\to-\infty$.
When considering instead a solid boundary at the bottom, as in the subsequent numerical analysis, this condition is replaced with a no-slip condition.

\subsection{Equilibrium state and linear theory}

We assume the system at rest as being represented with flat interfaces, i.e. $\eta(x,t)=\zeta(x,t)=0$.
The unperturbed hydrostatic component of the pressure field in $\Omega_1$ is thus recovered classically by integrating over the layer depth, reading
\begin{equation}
P_1(z) = - \alpha_f z .
\end{equation}

Similarly, integrating over the lower layer and assuming the pressure jump across the (flat) material interface (given by expression \rf{dbcn2}), it yields
\begin{equation}
P_2(z) = \alpha_{_f} + \alpha_{_m} - (z+1) .
\end{equation}

We consider the perturbed pressure solutions within both layers as departing from this equilibrium state, while considering disturbances in the form of monochromatic waves, i.e.
\begin{equation}
\label{pj}
p_j(x,z,t) = P_j(z) + \widehat{p}_j(z) \ue^{\ui(k x - \omega t)} ,
\end{equation}
where we assume the perturbation fields (denoted with a hat overscript) as having infinitesimal magnitudes.
The Fourier components are represented with a real spatial wavenumber $k$ and a complex frequency $\omega=\omega_r+\ui\omega_i$ ($\omega_r$ describes the oscillatory contribution whereas $\omega_i$ outlines the growth rate).

The base flow is not unique as it depends on the forcing term $f$ (no exact solution is known without external body force).
Hence, we assume a horizontal and monotonic velocity profile in the upper layer in the form of $\bm{u}_1=U(z)\bm{e}_x$ and we seek solutions fulfilling the boundary conditions \rf{kc1},\rf{kc2}, \rf{dbct1} and \rf{dbct2} with flat interfaces.
Obviously, this velocity profile should be characterised by an inflection point and the flow will thus be subject to inflectional instability due to Rayleigh--Fj{\o}rtoft criterion \citep{R80,F50}. 
Arbitrarily, we choose a third-order polynomial (although it is not the only choice) given in its simplest form (depicted in figure \ref{Fig1}) by
\begin{equation}
\label{Uz}
U(z) = - \pazocal{F} \left( 2z^3 + 3z^2 - 1 \right) .
\end{equation}
As an example, we highlight a second candidate solution, namely $U(z)=\pazocal{F}\sech^2{[z/(z+1)]}$, although we will restrict ourselves to \rf{Uz} in this study (this squared hyperbolic secant profile is used, for instance, in \cite{DT94} or in \cite{LH98}).

The perturbed velocity fields depart from this base state, as for the pressure, and read
\begin{equation}
\label{uj}
\bm{u}_1(x,z,t) = U(z) \bm{e}_x + \widehat{\bm{u}}_1(z) \ue^{\ui(k x - \omega t)} , \quad \bm{u}_2(x,z,t) = \widehat{\bm{u}}_2(z) \ue^{\ui(k x - \omega t)} .
\end{equation}

We follow the linear theory by substituting the perturbed fields \rf{pj} and \rf{uj} in \rf{eom1}--\rf{eom2} and by retaining only the terms linear in perturbations.
Then, we follow the work of \cite{KT94} to obtain the reduced perturbation equations.
This procedure consists in applying twice the curl operator on \rf{eom1} and projecting then  along the vertical.
Doing so, we eliminate the horizontal velocity component and the pressure field from the system.
Nevertheless, the boundary conditions require to implement the pressure contributions.
We therefore apply the horizontal divergence operator on the Navier--Stokes equations to express the perturbed pressure fields in terms of the vertical velocity perturbation.
We obtain
\begin{alignat}{2}
\widehat{p}_1(z) &= \ui\alpha_{_f}k^{-2} \left[ \omega - k U - \ui\pazocal{R}^{-1} (\partial_{zz} - k^2) \right] \partial_z\widehat{w}_1 + \ui\alpha_{_f}k^{-1} U' \widehat{w}_1 , \quad &&\textrm{in} \:\: \Omega_1 , \\
\widehat{p}_2(z) &= \ui k^{-2} \left[ \omega -\ui\pazocal{R}^{-1} (\partial_{zz} - k^2) \right] \partial_z\widehat{w}_2 , \quad &&\textrm{in} \:\: \Omega_2 ,
\end{alignat}
where the tilde denotes the derivative operation with respect to the $z$-coordinate.

At last, the linearised equations of motion within each one of the fluid layers reduce to
\begin{alignat}{2}
\label{leom1}
\left[ \omega - k U - \ui\pazocal{R}^{-1} (\partial_{zz} - k^2) \right] (\partial_{zz} - k^2) \widehat{w}_1 + k U'' \widehat{w}_1 &= 0 \quad &&\textrm{in} \:\: \Omega_1 , \\
\label{leom2}
\left[ \omega - \ui\pazocal{R}^{-1} (\partial_{zz} - k^2) \right] (\partial_{zz} - k^2) \widehat{w}_2 &= 0 \quad &&\textrm{in}  \:\: \Omega_2 ,
\end{alignat}
where the boundary conditions at the free surface are
\begin{align}
\label{lkc1}
\widehat{w}_1 + \ui (\omega - k \pazocal{F} ) \widehat{\eta} &= 0 , \\
\left[ \omega - k \pazocal{F} - \ui\pazocal{R}^{-1} (\partial_{zz} - 3k^2) \right] \partial_z\widehat{w}_1 + \ui k^2 ( 1 + k^2 \pazocal{B}_{\textrm{s}} ) \widehat{\eta} &= 0 , \label{bcsn} \\ 
(\partial_{zz} + k^2) \widehat{w}_1 &= 0 \label{bcst} ,
\end{align}
and the conditions at the material interface are
\begin{align}
\widehat{w}_1 + \ui \omega \widehat{\zeta} &= 0 , \label{bcm1} \\
\left[ \omega - \ui\pazocal{R}^{-1} (\partial_{zz} - 3k^2) \right] \left( \alpha_{_f} \partial_z\widehat{w}_1 - \partial_z\widehat{w}_2 \right) + \ui k^2 ( \alpha_{_m} \omega^2 - k^2 \pazocal{B}_{\textrm{b}} + \alpha_{_f} - 1 ) \widehat{\zeta} &= 0 , \label{bcm2} \\ 
\pazocal{R}^{-1} \left( \partial_{zz} + k^2 \right) \left( \alpha_{_f} \widehat{w}_1 - \widehat{w}_2 \right) - \alpha_{_m} k^2 \widehat{\zeta} &= 0 , \label{bcm3} \\
\widehat{w}_1 - \widehat{w}_2 = \partial_z\widehat{w}_1 - \partial_z\widehat{w}_2 &= 0 ,  \label{w1w2}
\end{align}
with $\widehat{w}_2 \to 0$ and $\partial_z\widehat{w}_2 \to 0$ at $z\to -\infty$ when considering the lower layer in the deep water limit.
The reduced expressions for the jump conditions in tangential stress are obtained by multiplying the original linearised conditions with $-\ui k$ and then using the continuity equation \rf{eom2} to eliminate the horizontal velocity disturbances.

For the sake of completeness, we give in Appendix \ref{appA} the inviscid counterpart of the latter set of equations and its reduction to the Nemtsov problem when the background flow is uniform \citep{N85,LK20,LK22}.

\subsection{Global stability from analytical and numerical approaches} \label{Sec23}

In the context of such problems, it is standard practice to express the linearised set of equations \rf{leom1}--\rf{w1w2} in the form of a polynomial eigenvalue problem (generally of second-order) for the eigenfrequency $\omega$ and the disturbance fields \citep{TM01}.
In classical mechanics, many finite-dimension holonomic systems derived from the Euler--Lagrange equations can be formulated in a similar form when linearised around an equilibrium state (cf. \cite{KM07} and references therein).
However, our dynamical system is characterised by an infinite number of degrees of freedom and incorporates dissipative and non-conservative forces.
Furthermore, the inviscid limit does not reduce to the viscous configuration as a result of frictional effects at the interfaces.
It is therefore not possible to interpret the following instability mechanisms as describing the departure of a relative equilibrium with Hamiltonian symmetry when dissipation is introduced.
In other words, the system under investigation does not fall within the paradigmatic concept of dissipation-induced instabilities \citep{BKMR94,Kirillov}. 

Using linear algebra, we express equations \rf{leom1}--\rf{w1w2} in the form of a quadratic eigenvalue problem
\begin{equation}
\label{qep}
\left( \omega^2 \pazocal{M} + \omega \pazocal{S} + \pazocal{K} \right) \widehat{\bm{\xi}} = \bm{0} ,
\end{equation}
where we eliminated the surface elevations from the state vector $\widehat{\bm{\xi}}=(\widehat{w}_1,\widehat{w}_2)^T$ by means of kinematic conditions \rf{lkc1} and \rf{bcm1}.
We emphasise, once again, that the differential operators $\pazocal{M}$, $\pazocal{S}$ and $\pazocal{K}$ do not necessarily correspond to mass, dissipation and stiffness matrices when discretised, as in the expository Chetaev systems \citep{BKMR94}.
Interestingly, we note that when $\alpha_m=0$ or when $\alpha_f=1$ and $\Bob=0$, the spectral problem \rf{qep} reduces to a linear eigenvalue problem in $\omega$.

In the rest of this section, we present the two procedures we use to determine the stability of our system at a fixed set of parameters $(k,\alpha_f,\alpha_m,\Fr,\Rey,\Bos,\Bob)$.

The first is to compute the exact form of the eigenfunctions $\widehat{w}_j(z)$ (up to arbitrary constants) and to substitute this ansatz within the boundary conditions.
Doing so allows to recover the dispersion relation by seeking for nontrivial solutions in the kernel of the associated linear system (i.e. by setting the determinant to zero).
However, as demonstrated by \cite{Chandra} when computing the stability of the viscous Rayleigh--Taylor problem, the characteristic equation results in a non-algebraic expression.
This observation prevents the benefits of using classical polynomial stability criteria, e.g. the Routh--Hurwitz or the Li\'enard--Chipart criterion \citep{LC14,Kirillov}.
We still demonstrate the efficiency of analytical dispersion relation by recovering numerically the solutions via some root-finding algorithms.
Noteworthy, in the inviscid limit and for a material interface of finite chord length, the Nemtsov problem reduces to a non-algebraic integro-differential dispersion relation \citep{N85,LK22}.

A second way of determining the global stability of our system is by solving numerically the boundary value problem \rf{leom1}--\rf{w1w2} by means of a pseudo-spectral collocation method \citep{Boyd}.
The basic idea is to expand the eigenfields on a high-degree orthogonal polynomial basis and to evaluate this quadrature at given collocation points .
We choose in the present case to consider the Chebyshev polynomials truncated at the $(N+2)$-th order and evaluated on the Chebyshev grid $z_i=\cos{(i\pi/N)}$, $i=1,\dots,N-1$.
This method is notorious to exhibit spectral convergence when varying the truncation order $N$, although the accuracy may worsen as this number becomes too large due to the conditioning of the linear system \citep{GO77}. 
Indeed, the equations of motion within the layers are linear in $\omega$ and hence, the operator $\pazocal{M}$ is singular (it only contains zero entries except for a few rows).
For each computations, we verify the accuracy of the numerical method by a convergence analysis of the spectrum, by discarding the spurious eigenvalues and by comparing with analytical results when available.
We use in practice a truncation order of $N=32$ for the motionless case with $\Fr=0$ and $N=64$ when considering a parallel base flow.
For numerical reasons, we decide to consider a solid wall underneath the lower layer at a location $z=-H$ (only when computing the boundary value problem).
The evanescence conditions are thus replaced by no-slip conditions, although assuming $H\gg1$ supposedly coincides with the deep water regime.

\section{Stability analysis of the motionless configuration} \label{sec2}

We treat here the case where the background flow is absent from the system, i.e. $\Fr=0$.
Under such assumption, the configuration is reminiscent of the Rayleigh--Taylor problem of two immiscible fluid layers (viscous or not) separated by a liquid-liquid interface.
The classical analysis yields the celebrated stability criterion $\alpha_f>1$ for the onset of linear instability due to the potential energy being transferred from the heavier to the lighter fluid in the same direction as gravity applies \citep{T50}.

We demonstrate a new instability by divergence within the Rayleigh--Taylor stability domain (that is for $\alpha_f<1$) induced by the wave-interaction and the discontinuity in tangential stress at the material interface.
First, we prove the unconditional stability of a series of non-interacting configurations.
Then, we present the exact dispersion relation of the two-phase system and we give evidence for this new instability mechanism by means of numerical and analytical results.

\subsection{Dispersion relations of the uncoupled system}

We present the stability analysis of two specific configurations: a viscous fluid layer with a free surface and two semi-infinite fluid layers enclosing a material interface.

\subsubsection{The case of a viscous fluid layer with a free surface}

Our system reduces to a single motionless layer by assuming the interface as being massless (i.e. $\alpha_m=0$) and the two fluid layers as having the same density (i.e. $\alpha_f=1$).
In this case, the eigenfunction $\widehat{w}_1$ is given as the solution of equation \rf{leom1} with $\Fr=0$, which reads
\begin{equation}
\label{w1}
\widehat{w}_1(z) = A_1 \ue^{k z} + B_1 \ue^{-k z} + C_1 \ue^{q z}  + D_1 \ue^{-q z} ,
\end{equation}
where capital letters denote arbitrary constants and where
\begin{equation}
\label{q}
q = k \sqrt{1 - \ui\omega\Rey/k^2} = k \sqrt{\Omega + 1} ,
\end{equation}
with the convention that $q$ is taken with a positive real part \citep{Chandra}.

Taking into consideration the evanescence condition from the deep water assumption, the eigenfield \rf{w1} simplifies to
\begin{equation}
\widehat{w}_1(z) = A_1 \ue^{k z} + C_1 \ue^{k \sqrt{\Omega+1} z}  .
\end{equation}

Substituting the latter expression within the boundary conditions \rf{bcsn}--\rf{bcst} at the surface, it yields a linear system for the scalar constants.
We obtain a solvability condition by setting its determinant to zero, yielding the dispersion relation of a motionless viscous deep water layer with a free surface 
\begin{equation}
\label{dr0}
\pazocal{D}_{\textrm{dw}}(\Omega,k) = 1 + k^2\Bos + k^3\Rey^2 \left[ (\Omega+2)^2 - 4\sqrt{\Omega+1} \right] 
= \pazocal{D}_{0}(\Omega,k) + k^2 \Bos = 0  ,
\end{equation}
where we recall the relation $\omega=\ui k^2 \Omega/\Rey$ to recover the original eigenfrequency.
The system is said to be unconditionally stable against infinitesimal perturbations if and only if every roots observe the condition $\Re{(\Omega)}<0$.
Proving this statement requires first to convert the dispersion relation \rf{dr0} into an algebraic expression for $\Omega$ (by isolating the square root and squaring both sides) and then, to apply the Li\'enard--Chipart polynomial criterion \citep{LC14} to determine the root locations within the complex plane.
We notice that the leading principal minors of the Hurwitz matrix are always positive for all values of $(k,\Rey,\Bos)$ and thus, the system governed by this dispersion relation is always stable (every zeros have negative real parts).

For the sake of completeness, we give the dispersion relation when a solid wall is present at $z=-1$.
Introducing the classical change of variable $\Omega=X^2-1$, the characteristic equation governing the stability of a viscous free surface layer at finite-depth reduces to
\begin{equation}
\label{drfd}
\pazocal{D}_{\textrm{fd}}(X,k) = \pazocal{D}_{\textrm{w}}(X,k) - \pazocal{D}_{\textrm{w}}(-X,k) + \pazocal{D}_{\textrm{w}}(-X,-k) - \pazocal{D}_{\textrm{w}}(X,-k) = 0  ,
\end{equation}
where $\pazocal{D}_{\textrm{w}}(X,k) = \pazocal{D}_{1}(X,k) - 4\mu_1(X,k) - 4X\mu_2(X,k)$, and
\begin{gather}
\pazocal{D}_{1}(X,k) = \pazocal{D}_{\textrm{dw}}(X,k) \ue^{k(X+1)} , \label{dr1} \\
\mu_1(X,k)=k^3\Rey^{-2}(X^2-1)^2, \quad 
\mu_2(X,k)=k^3\Rey^{-2}(X^2+1) . \label{mu}
\end{gather}
As in the deep water configuration, we can show that \rf{drfd} is unconditionally stable against infinitesimal disturbances.

\subsubsection{The case of a material interface enclosed by two viscous fluids}

It is well-known that two superposed immiscible fluid layers with distinct (or not) densities are subject to the Rayleigh--Taylor instability whenever the lighter fluid is accelerated in the direction of the heavier fluid \citep{R82,T50}.
This phenomenon is due to gravity and yields the growth of disturbances at the interface, generating typical finger-like structures \citep{L50}.
Removing the free surface from our physical system and considering the fluid above as occupying the whole upper-space, we obtain a configuration similar to the original Rayleigh--Taylor model, although the interface displays inertia effects.
We propose to investigate the stability of this uncoupled system in the region with stable potential energy, i.e. for $\alpha_f<1$.

After application of the evanescence conditions at $z\to\pm\infty$, the eigenfields in their respective domains are given by
\begin{equation}
\widehat{w}_1(z) = B_1 \ue^{-kz} + D_1 \ue^{-q z}, \quad \widehat{w}_2(z) = A_2 \ue^{kz} + C_2 \ue^{q z},
\end{equation}
where $q$ is still given by \rf{q}.
We express the eigenfunction in the lower region in terms of the constants in the upper domain from the condition of velocity continuity \rf{w1w2} across the interface.
It yields
\begin{equation}
A_2 = -\frac{B_1(k+q) + 2D_1q}{k-q}, \quad C_2 = \frac{D_1(k+q) + 2B_1q}{k-q},
\end{equation}
where we note that the case $\omega=0$ is singular (but not of interest here).

We substitute the later solutions within the boundary conditions \rf{bcm1}--\rf{bcm3} to recover the governing dispersion relation of this modified Rayleigh--Taylor system.
Noteworthy, the boundary conditions are written here at $z=0$ instead of $z=-1$ without loss of generality.
We obtain after simplification
\begin{align}
\label{drm}
\pazocal{D}_m(\Omega,k) = &\left[ k^3\Rey^{-2} (\alpha_f+1)(\alpha_f+1+k\alpha_m) \Omega^2 + 4k^3\Rey^{-2} (\alpha_f^2+1) \Omega + k^2 \Bob (\alpha_f+1) \right. \nonumber \\
&\left. - (\alpha_f-1)(\alpha_f+1+k\alpha_m) \right] ( 1 + \sqrt{\Omega+1} ) + 4k^3\Rey^{-2} \left[ \alpha_f \Omega + (\alpha_f-1)^2 \right] \Omega \nonumber \\
&-2k\alpha_m (\alpha_f-1) .
\end{align}

The characteristic equation \rf{drm} is obviously non-algebraic in the eigenfrequency, prohibiting the application of polynomial methods to investigate its stability analysis.
However, as demonstrated in Appendix \ref{appB}, we can prove the unconditional stability of the system in the regime $\alpha_f<1$ by means of complex analysis and successive applications of Rouch\'e's theorem \citep{R62}.
We note that for a massless interface ($\alpha_m=0$), expression \rf{drm} reduces to the dispersion relation of two superposed fluid layers of distinct densities but same viscosity derived, e.g., in \citet[\S X,~pg.~444]{Chandra}.
When we neglect the effects induced by surface tension ($\Bob=0$) and we assume two liquids of the same density ($\alpha_f=1$) we find the explicit solution $\Omega=-4(1+k\alpha_m)/(2+k\alpha_m)^2$ that is stable.

\subsection{A new instability mechanism induced by viscous fluid-structure interaction}

We have shown in the previous two subsections that the waves emitted by the flexible boundaries, when decoupled, do not trigger instability as soon as $\alpha_f<1$.
In this section, we demonstrate the onset of instability when considering the system with nonconservative tangential forces induced by the fluid load and the interaction with the free surface.
Furthermore, we demonstrate the existence of this new mechanism within the stability domain of Rayleigh--Taylor instability, i.e. for $\alpha_f<1$.

Let consider the general solutions of \rf{leom1}--\rf{leom2} in their respective domains, i.e.
\begin{equation}
\widehat{w}_1(z) = A_1 \ue^{kz} + B_1 \ue^{-k z} + C_1 \ue^{qz} + D_1 \ue^{-q z}, \quad \widehat{w}_2(z) = A_2 \ue^{kz} + C_2 \ue^{qz} ,
\end{equation}
while assuming the lower layer as being semi-infinite.
Substituting these solutions within the set of boundary conditions \rf{lkc1}--\rf{w1w2} yields a $8\times8$ linear system whose solvability condition yields the characteristic equation.
After some algebraic manipulations and using once again the change of variable $X=\sqrt{1 - \ui\omega\Rey/k^2}$ (where the square root branch is chosen with positive real part), we finally obtain the dispersion relation of the coupled system as
\begin{equation}
\label{drfull}
\pazocal{D}(X,k) = \gamma_1 \pazocal{D}_1(X,k) + \gamma_2 \pazocal{D}_1(-X,k) + \gamma_3 \pazocal{D}_1(X,-k) + \gamma_4 \pazocal{D}_1(-X,-k) + \gamma_5 = 0 ,
\end{equation}
where $\pazocal{D}_1$ is retrieved from \rf{dr1} and with the $\gamma_j(X,k)$ coefficients being given in Appendix \ref{appC}.

Expression \rf{drfull} is the main result of this section and will be used to compute the growth rate of this new linear instability at a fixed set of parameters.

\subsection{Comparison analysis between numerical and analytical results}

\begin{figure}[t!]
    \centering
    \includegraphics[width=\textwidth]{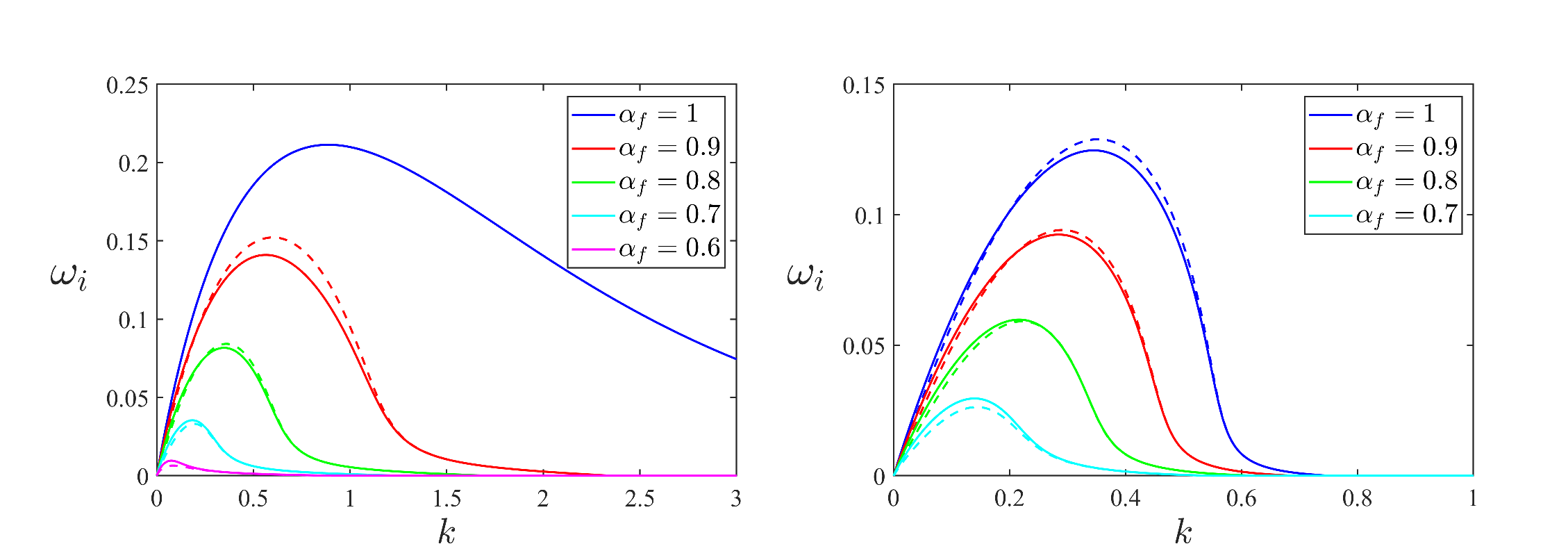}
    \caption{Growth rates of the motionless system computed from the deep water dispersion relation \rf{drfull} (solid lines) and the finite-size boundary value problem \rf{leom1}--\rf{w1w2} (dashed lines) at the fixed parameters $\alpha_m=1$, $\Rey=10^{3}$ and $H=10$. 
    The left panel corresponds to the case without surface tension $\Bob=\Bos=0$ whereas the right panel corresponds to $\Bob=1$ and $\Bos=1/3$.}
    \label{Fig2}
\end{figure}

This section is dedicated to the computation of eigenfrequencies of the coupled system, either by solving the boundary value problem \rf{leom1}--\rf{w1w2} with no-slip condition at the bottom or by solving expression \rf{drfull} in deep water.
In the former case, the pseudo-spectral collocation method described in section \ref{Sec23} is employed, whereas in the latter case, a classical root-finding algorithm is utilised, for instance, the function \textsf{fsolve} implemented in \textsc{Matlab}.
The resulting growth rates are presented in Figure \ref{Fig2}, with solid and dashed lines representing the solutions of the dispersion relation and the boundary value problem, respectively.
It is noteworthy that in the case where $\alpha_f=1$ and when surface tension is absent, the two curves are indistinguishable and the two results are in perfect agreement (this case is simpler to compute as it reduces to a linear eigenvalue problem, cf. section \ref{Sec23}).
The discrepancies between the other curves can be attributed to the bottom condition, namely deep water versus finite-depth.
When the density ratio $\alpha_f$ is not sufficiently large, no instability is observed, but evidence of a positive growth rate emerging from the short wavelength regime at a finite threshold is present.
A numerical investigation revealed that the threshold for this bifurcation is $\alpha_f\approx0.44$ in the configuration without surface tension (left panel).

As the value of $\alpha_f$ increases, the region of instability broadens and the growth rate increases in magnitude.
When introducing surface tension into the system (as illustrated in the right panel of figure \ref{Fig2}), we observe a damping effect of the instability in the short wavelength limit.
This observation is in agreement with the analysis of Rayleigh--Taylor mechanism when capillary effects are taken into account \citep{Chandra}.
Allowing the viscosity and the structural property of the membrane to vary (through the Reynolds number $\pazocal{R}$ and the mass ratio $\alpha_m$) primarily affects the magnitude of the growth rate, but not the nature of the unstable mode.

\begin{figure}[t!]
    \centering
    \includegraphics[width=\textwidth]{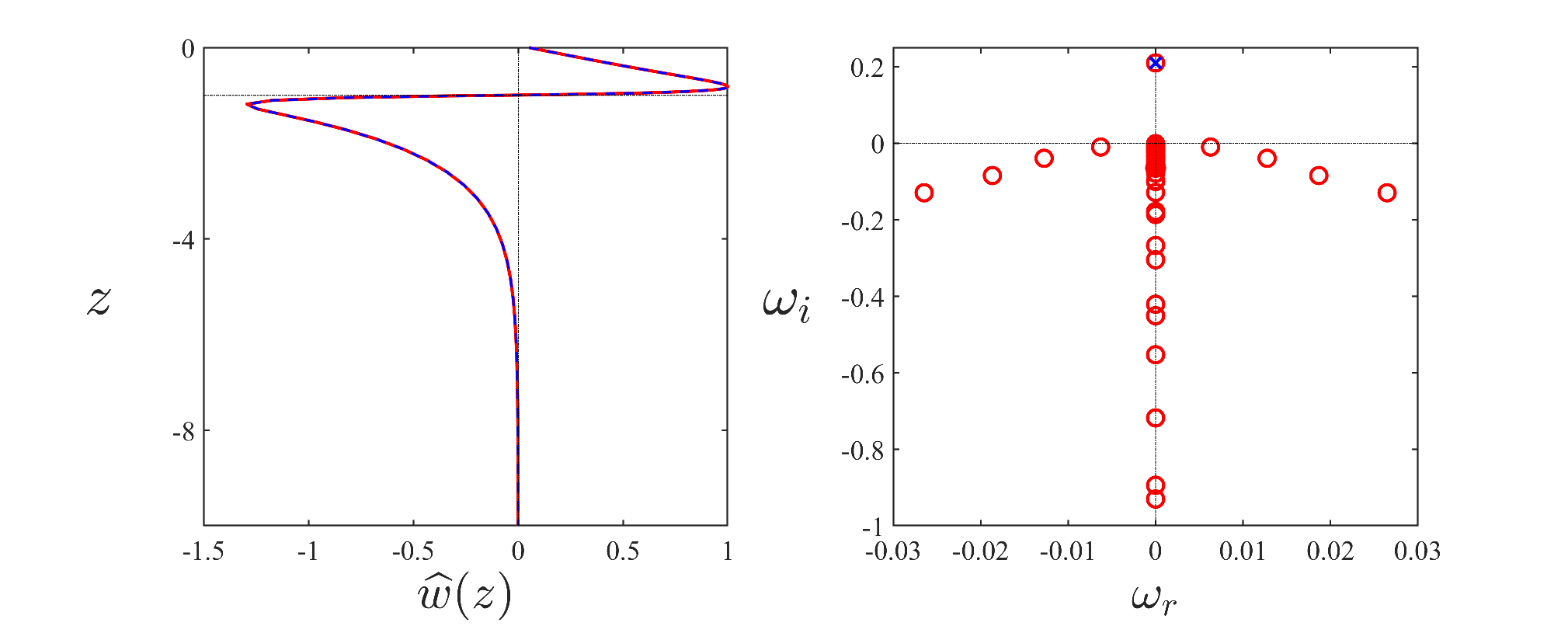}
    \caption{Eigenmode and spectrum computed from the configuration displayed in figure \ref{Fig2} with $k=1$, $\alpha_f=1$ and without surface tension. 
    The blue cross on the right panel depicts the most unstable eigenvalue computed from \rf{drfull} for which we represented the associated eigenfunction on the left panel. 
    The analytical and numerical eigenfunctions are represented in dashed blue line and solid red line, respectively.}
    \label{Fig3}
\end{figure}

It is noteworthy that the real components of the eigenfrequencies are identically zero.
This mechanism is therefore of the same nature as the divergence instability of a thin membrane immersed in an inviscid fluid flow \citep{MA21}.
This observation is made evident when looking at the numerical spectrum of the system for $k=1$ and $\alpha_f=1$, as represented in the right panel of figure \ref{Fig3}.
It can be observed that the spectrum contains a purely imaginary eigenvalue with a positive imaginary part, which corresponds to this new instability mechanism.
The associated eigenfunction, represented in the left panel, exhibits a perfect agreement between the analytical solution (dashed blue line) and the numerical solution (continuous red line).
We see that the vertical velocity is varying the most when being closer to the interface.
We interpret this as a consequence of the membrane not being flat any longer, because of the onset of exponential growth.

In conclusion, we demonstrated evidence for a new instability phenomenon due to the interaction between a material interface and a free surface.
This result shares some similarities with the analysis done by \cite{N85} (described in Appendix \ref{appA}) although the predicted flutter phenomenon is inhibited at $\Fr=0$ (and is oscillatory by definition). 
Surprisingly, this mechanism triggers even when the fluid above is lighter than the fluid below or, equivalently, when the Rayleigh--Taylor instability is prohibited.
For this novel result to hold, it is imperative to consider the material property of the membrane ($\alpha_m\neq 0$) and to acknowledge the existence of the free surface.

\section{Membrane flutter in the presence of a parallel shear flow} \label{sec3}

\begin{figure}[t!]
    \centering
    \includegraphics[width=.33\textwidth]{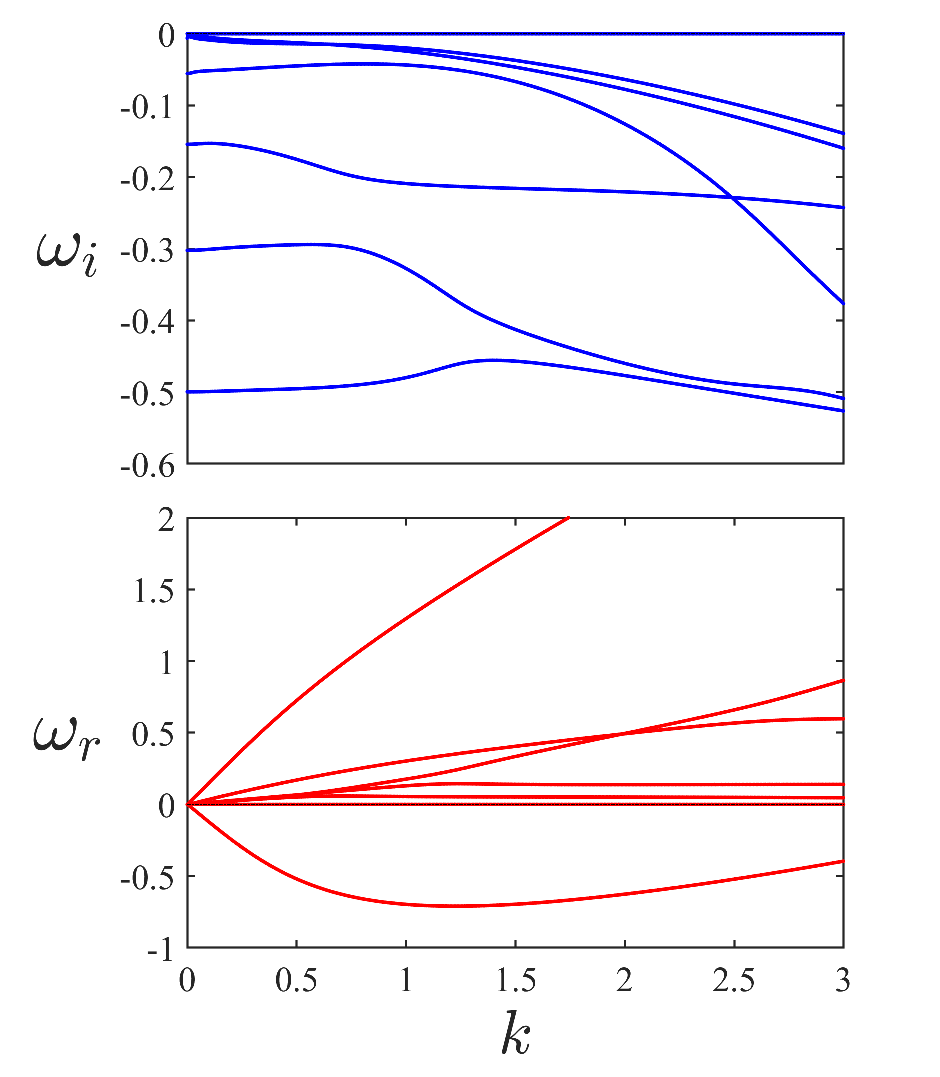}~
    \includegraphics[width=.33\textwidth]{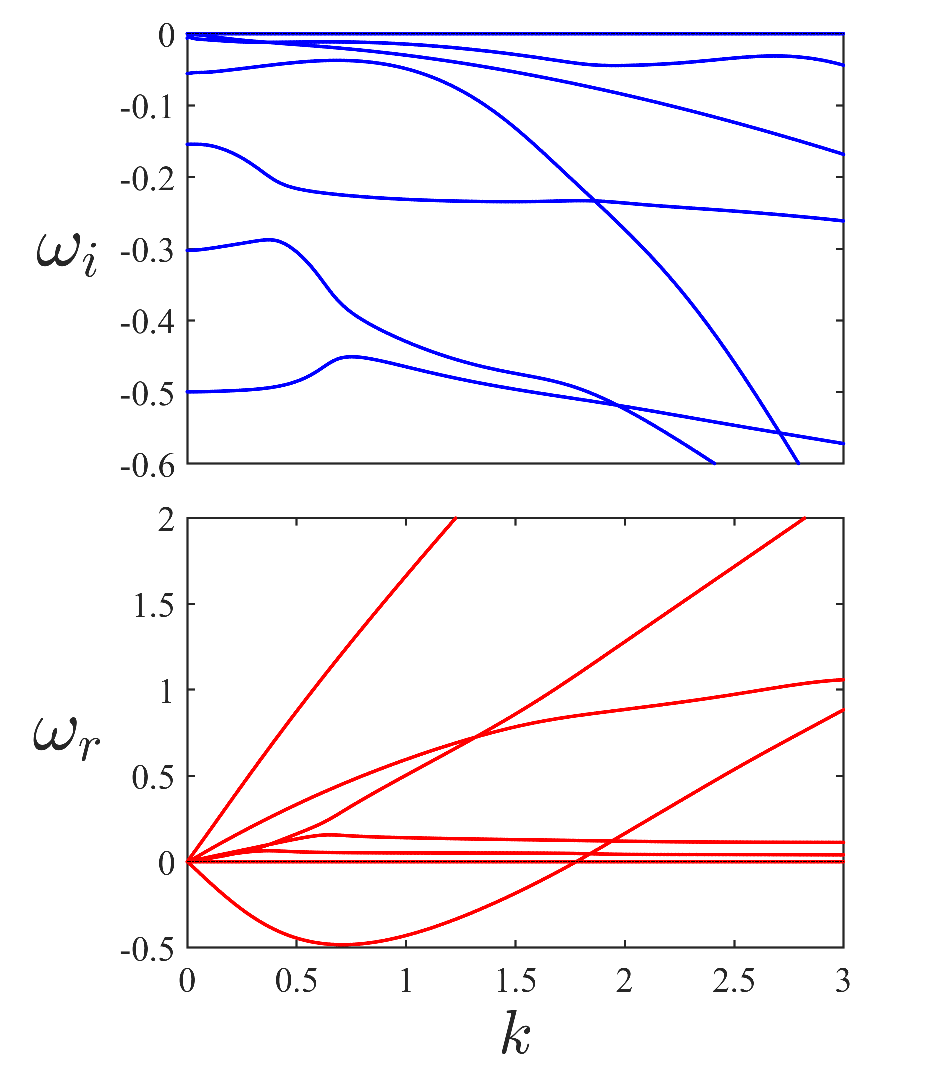}~
    \includegraphics[width=.33\textwidth]{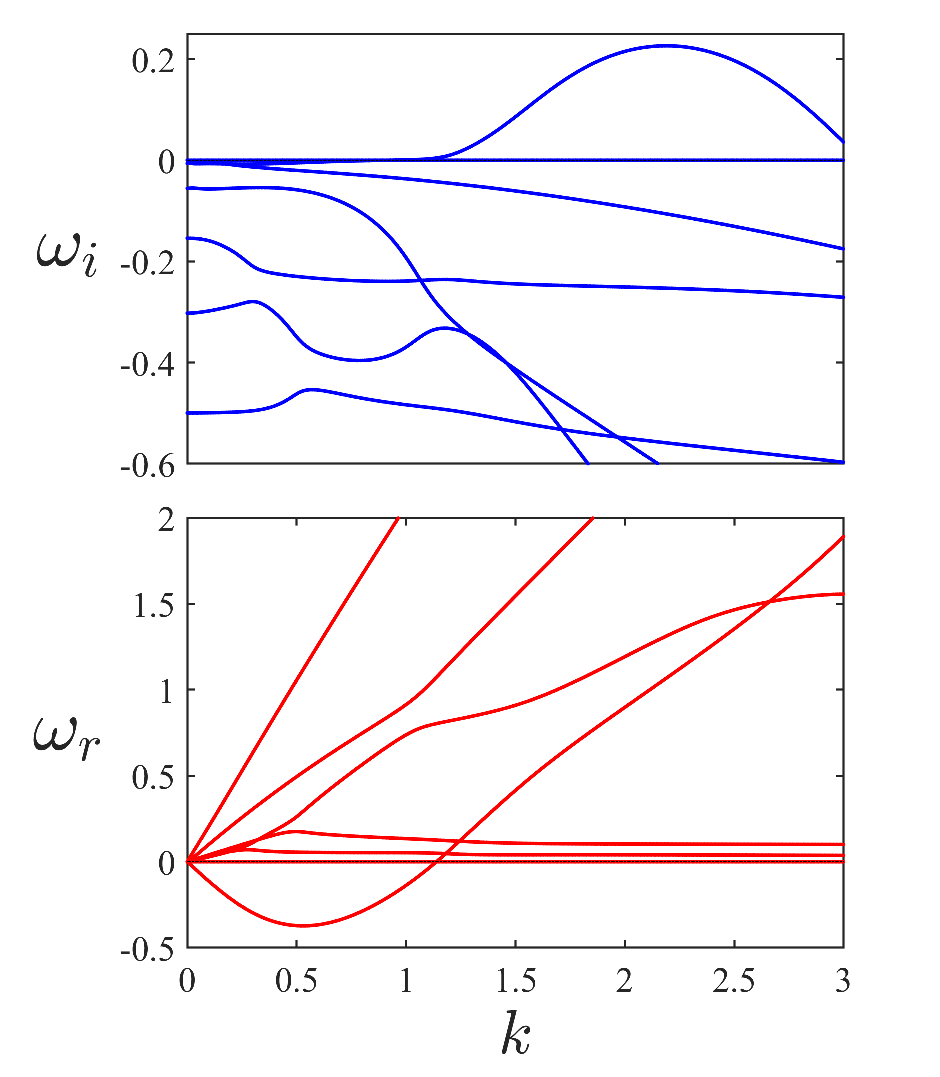}\\
    \includegraphics[width=.33\textwidth]{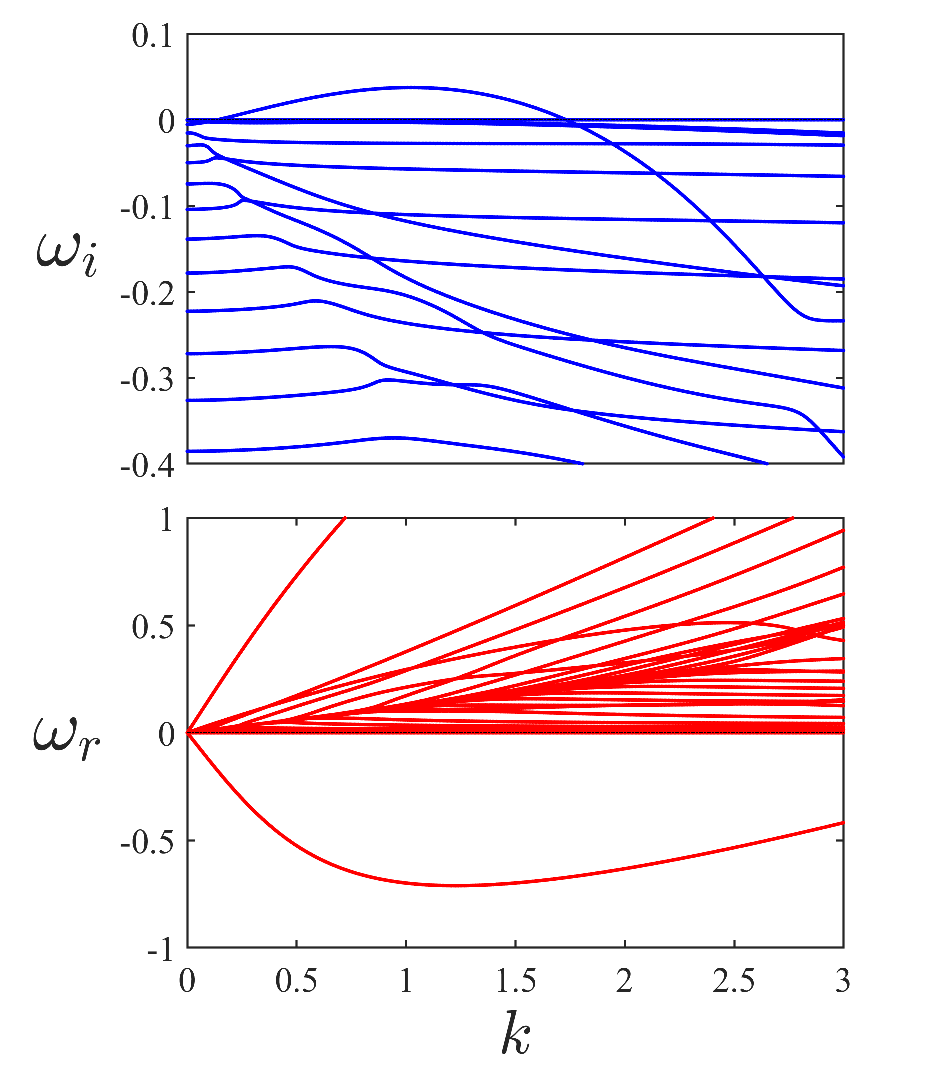}~
    \includegraphics[width=.33\textwidth]{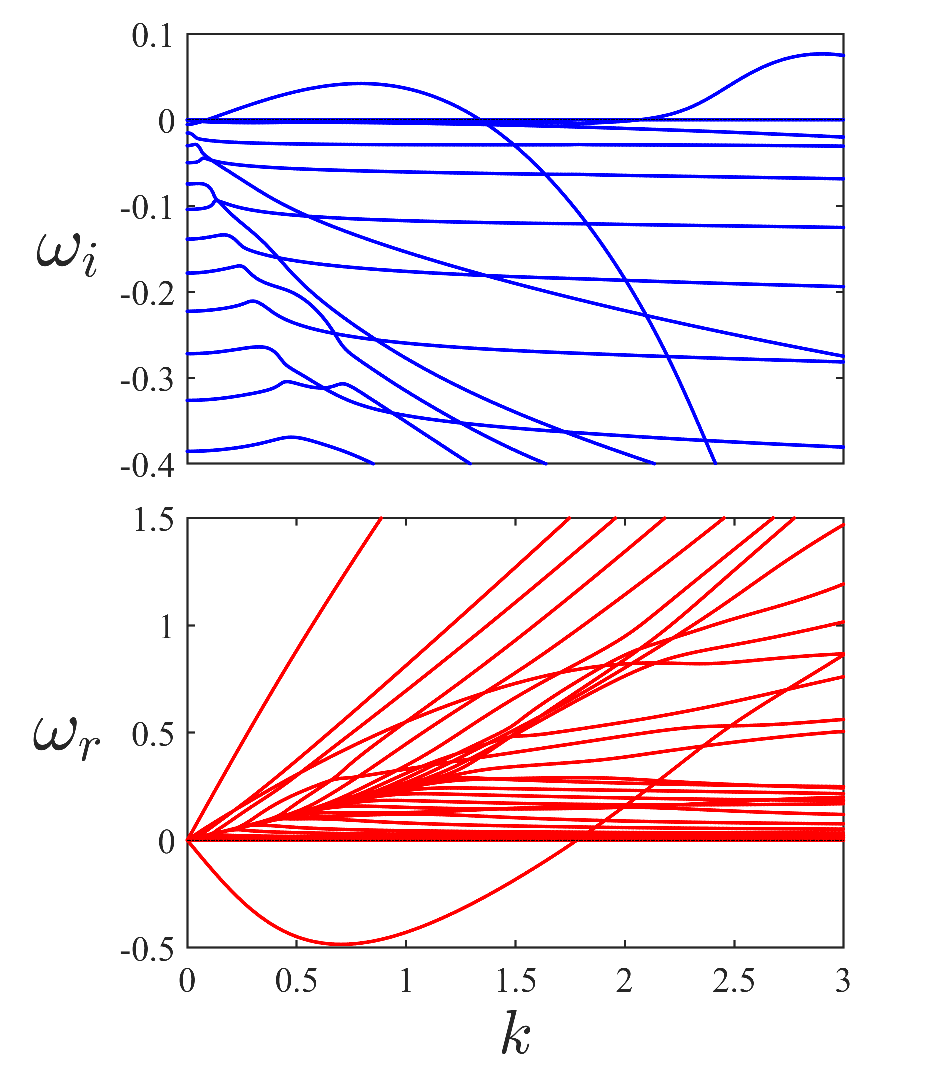}~
    \includegraphics[width=.33\textwidth]{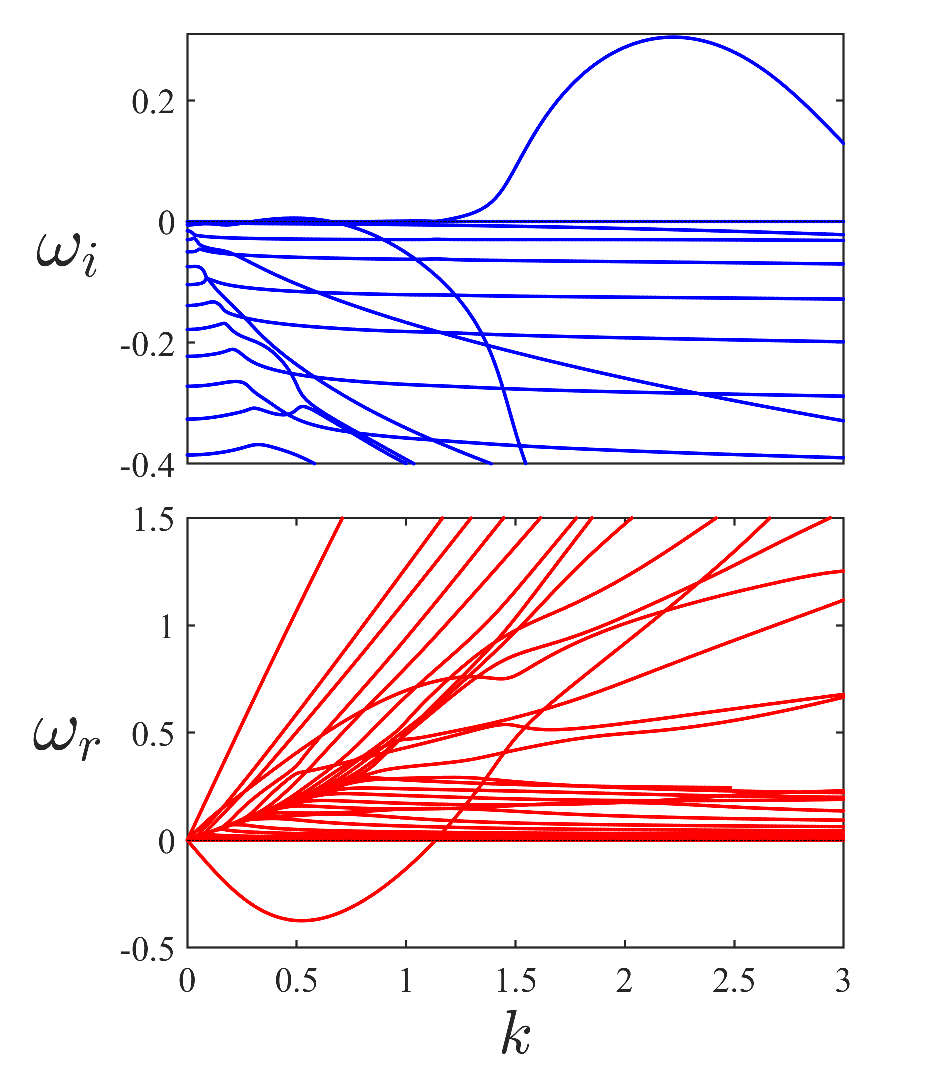}
    \caption{Dispersion curves for the system with a massless interface computed at $\alpha_f=1$, $\Bob=\Bos=0$, $\Rey=10^2$ (upper row) and $\Rey=10^3$ (lower row) in a finite-depth domain of size $H=2$. 
    The panels display the leading eigenvalues and, from left to right, $\Fr=0.5$, $\Fr=1$ and $\Fr=1.5$.}.
    \label{Fig4}
\end{figure}

Having established the existence of a new `static' instability for the motionless layers, we proceed to extend our analysis to the case where the upper layer is in parallel motion due to an external forcing.
We assume in the following a velocity profile of the form \rf{Uz} with the Froude number $\Fr$ as the control parameter.
In contrast with the previous section, computing the analytical solution of \rf{leom1} is not possible in general.
As a result, we rely extensively on the numerical solution of the boundary value problem to determine the global stability of the system.

We distinguish two configurations of interest in order to isolate the physical phenomena responsible for the out-of-equilibrium dynamics.
The first configuration corresponds to the massless interface and yields the growth of disturbances due to the inflection point in the background velocity profile \citep{R80,F50} or to the interaction of surface waves with the shear motion \citep{DT94,LH98}.
The second case is that of the coupled system without surface tension, which may be reduced to a linear eigenvalue problem (cf. section \ref{Sec23}).
The addition of friction at the interface allows the system to fall within the scope of the boundary-layer stability theory \citep{B60}.
This configuration allows for the growth of Tollmien--Schlichting waves at large Reynolds number \citep{B96} although it is also unstable due to the inflectional velocity profile.
Moreover, the system can also destabilised due to an over-reflectional process \citep{B99} as attested in the ideal case \citep{N85,LK20,LK22}.
Noteworthy, we fix in both cases the density ratio at $\alpha_f=1$ (no density stratification) such that the Kelvin--Helmholtz and the Rayleigh--Taylor instabilities are both prohibited.

\subsection{Shear-induced instability of a massless interface}

When the mass ratio of the interface is considered to be negligible (or, equivalently, when $\alpha_m$ is set to zero), the dynamical system simplifies to a three-layer flow with an upper gas-liquid interface and a lower liquid-liquid interface.
Furthermore, in the absence of density stratification, the motionless lower layer does not influence the global dynamics (due to the absence of surface tension), and may be regarded as a deep water condition.
This configuration is reminiscent of the inviscid model studied by \cite{LH98}.
In this case, the principal source of energy is generated by the shear motion, which triggers instability in the inviscid case due to the inflection point in the background flow \citep{R80,F50} or as a result of interaction with the free surface \citep{DT94}.

Figure \ref{Fig4} illustrates a series of dispersion curves for the leading eigenvalues (i.e. with the largest imaginary parts), computed at different Reynolds and Froude numbers.
In the case with $\Rey=10^2$, an unstable short-wavelength mode emerges when a finite threshold in Froude number is crossed.
This mode can be attributed to either the viscous counterpart of the inflectional instability or the shear-induced instability described in \citep{DT94}. 
At lower Froude numbers, these modes are damped by viscosity.

As the Reynolds number is increased up to $\Rey=10^3$, we notice a long-wavelength branch of instability at low Froude number.
This mode can be related to the growth of Tollmien--Schlichting waves due to the tangential stress at the flexible solid interface, in a manner analogous to the theory of boundary-layer stability \citep{B60,B96}.
As previously observed, the system still exhibits the same unstable mode at large Froude numbers, with both instabilities coexisting simultaneously.

It is important to note that the predictions of inflectional and shear-induced mechanisms are originally established in the ideal flow setting. 
In this study, we demonstrate the interplay between the viscous counterpart of these modes of instability and how they can be inhibited in certain limits.
The subsequent configuration incorporates the material property of the membrane to investigate the manner in which it yields the radiation-induced instability predicted by \cite{N85} and \cite{LK20,LK22}.

\begin{figure}[t!]
    \centering
    \includegraphics[width=.33\textwidth]{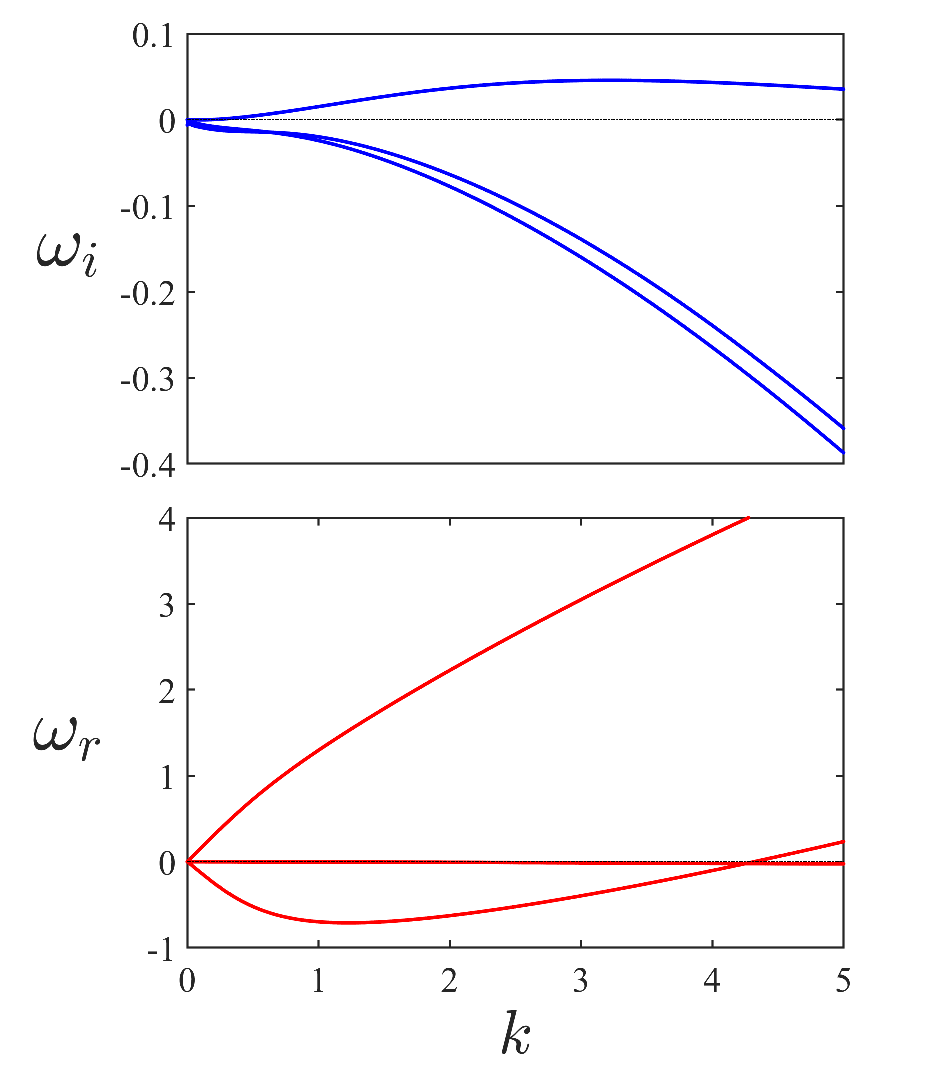}~
    \includegraphics[width=.33\textwidth]{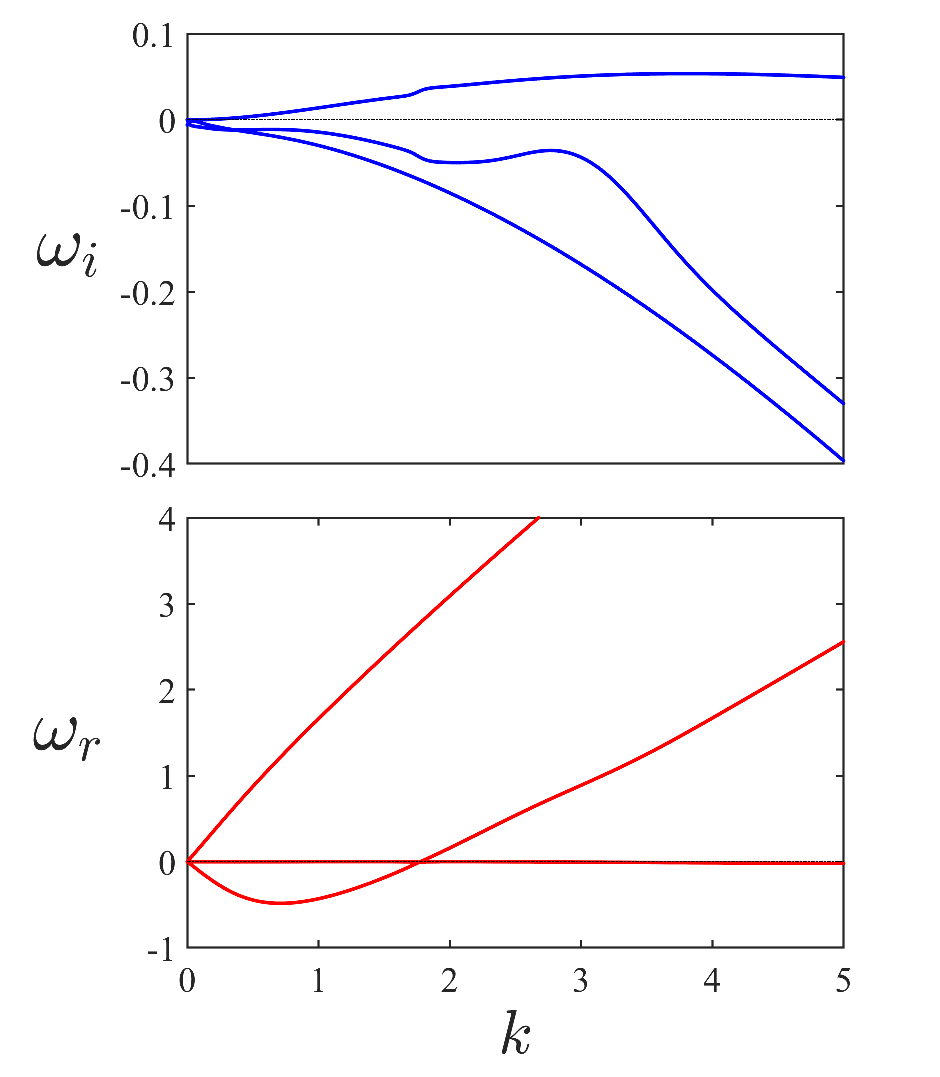}~
    \includegraphics[width=.33\textwidth]{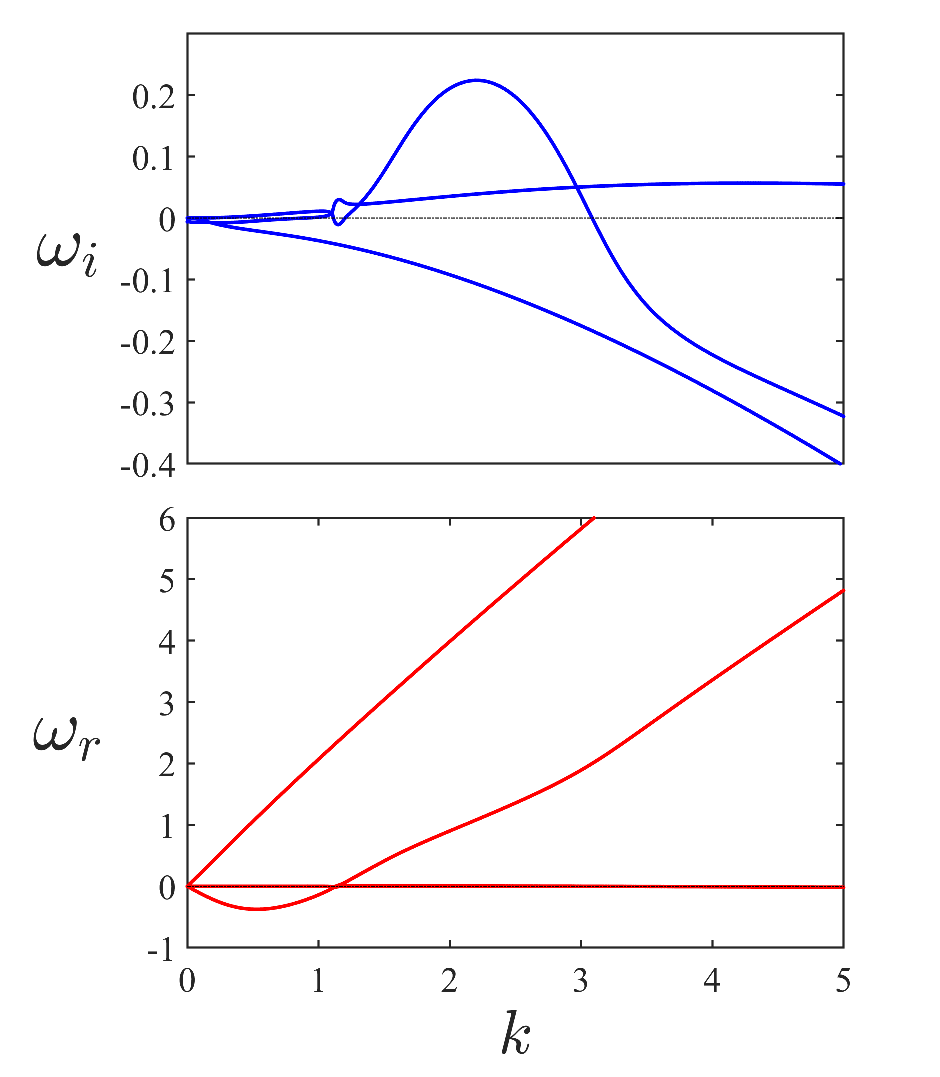}
    \caption{Same graphics as in figure \ref{Fig4} but for a material interface with mass ratio $\alpha_m=0.01$ and $\Rey=10^2$. 
    Are displayed only the 3 leading eigenvalues.}
    \label{Fig5}
\end{figure}

\begin{figure}[t!]
    \centering
    \includegraphics[width=\textwidth]{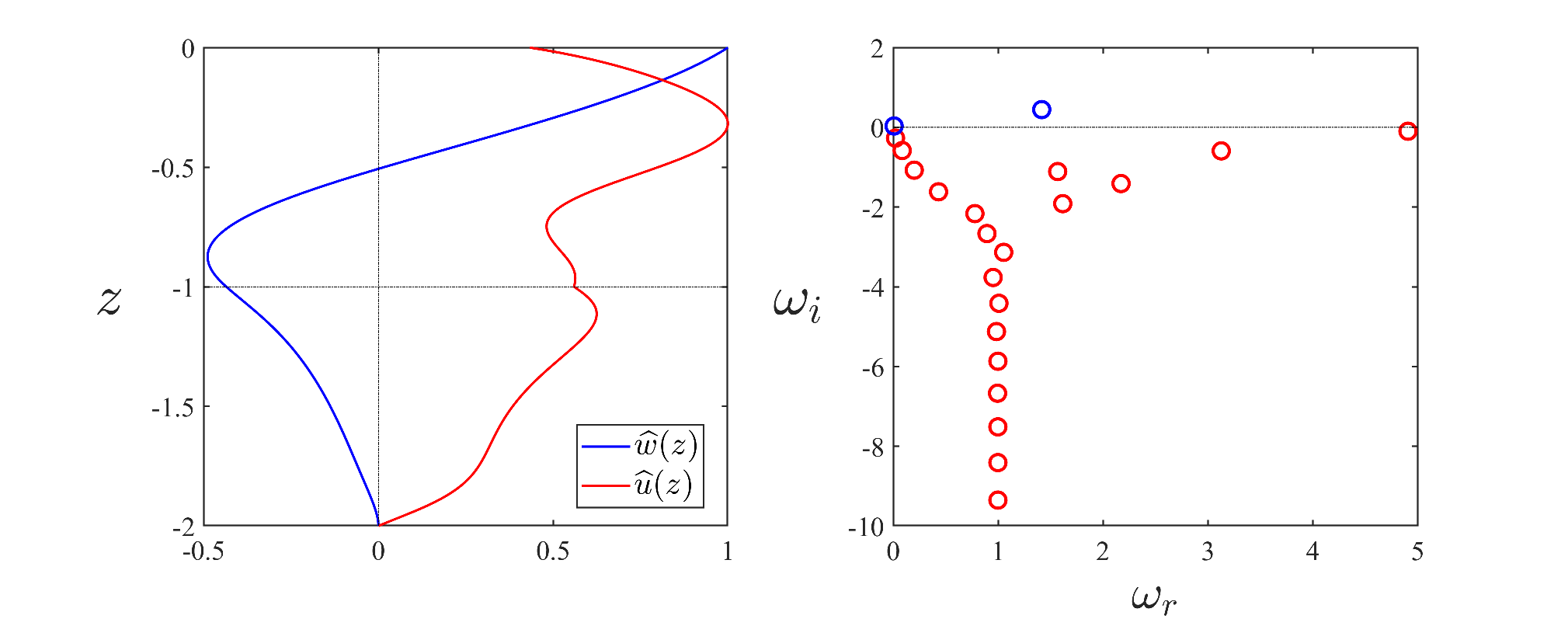}
    \caption{Numerical eigenmodes and spectrum computed at $k=2$ for the same parameters as in figure \ref{Fig5}, except that $\Fr=2$. 
    The blue circles denote eigenvalues with positive imaginary parts for which we represented the eigenfunctions associated to the mode with the largest growth rate.}
    \label{Fig6}
\end{figure}

\subsection{Radiation-induced instability of a material interface}

We now return to the configuration of a material interface with a prescribed mass, while still neglecting capillary effects.

We present in figure \ref{Fig5} the numerical dispersion curves for the configuration at $\alpha_m=0.01$, $\Rey=10^2$ and different Froude numbers as in figure \ref{Fig4} (except we only display the 3 leading eigenvalues).
It is evident that a new unstable mode emerges at low Froude number, spanning the entire range of horizontal wavenumbers.
This mode corresponds to the radiation-induced instability of surface and interfacial waves, which yields the membrane flutter predicted by \cite{N85} in shallow water and recently confirmed by \cite{LK20,LK22} in the finite-depth regime.
Noteworthy, this instability is connected to the radiation of negative energy waves within the region of the anomalous Doppler effect \citep{N76}.
As the magnitude of the background flow is increased, a second unstable mode emerges.
This mode is attributed to the background velocity profile and the eventual growth of Tollmien--Schlichting waves.
As illustrated in the right panel of Figure 5, this instability becomes dominant over a finite interval in $k$, while being damped in the short-wavelength regimes.

To highlight this behaviour, we display in figure \ref{Fig6} the discrete spectrum of the system and the associated eigenmodes for a configuration at $\Fr=2$ where the two instabilities are in competition.
It seems reasonable to suggest that the destabilisation mechanism in this context is due to the interplay between the inviscid inflectional theorem, the shear-induced branches and the boundary-layer stability.
We support this claim by noticing the similarity with the Orr--Sommerfeld spectrum of wall-bounded shear flows \citep{O07}.

\section{Discussion} \label{sec4}

This paper introduced unprecedented results on the stability of a viscous two-layer system with a material interface and a free surface.
We benefited from the use of analytical and numerical methods separately to compute the associated dispersion relation and the spectrum of the linearised operator, respectively.
The former procedure required to identify the analytic form of the solutions within the bulk whereas the latter used a discretisation of the boundary value problem with a pseudo-spectral collocation method.

As a preliminary investigation, we examined the scenario where both layers were motionless.
By considering a series of uncoupled systems, we demonstrated that global stability is maintained whenever the two interfaces are not interacting and when the potential forces are acting from the lighter to the heavier fluid.
When the system is coupled, both analytical and numerical procedures revealed the existence of a new instability (similar to the membrane divergence, cf. \cite{MA21}) in the parameter space that is stable against the Rayleigh--Taylor mechanism.
Furthermore, it was demonstrated that this mode persisted in the presence of surface tension and irrespectively of the viscosity magnitude.
This novel instability was interpreted as being induced by the non-conservative tangential forces and the fluid-structure interaction.

Subsequently, we introduced a background velocity flow within the upper layer.
This monotonic shear flow inevitably displays an inflection point (to fulfil the boundary conditions) and thus falls within the scope of Rayleigh--Fj{\o}rtoft stability criterion \citep{R80,F50}.
However, additional instability mechanisms also arise when the upper layer is in parallel motion and a competition between these different modes is to be expected.
As the eigenfunctions were not analytically tractable, we relied on solving the boundary value problem numerically to compute the dispersion curves of the system.
Doing so, we demonstrated the existence of the inflection-induced mode, the shear-induced mode (referred to as Branch I and Branch II from the inviscid theory of \cite{DT94}), the viscous mode (due to Tollmien--Schlichting waves as predicted by \cite{B60}) and the radiation-induced mode (due to the emission of surface gravity waves within the region of anomalous Doppler effect, see \cite{N85}).
We outlined the interplay between the different instability mechanisms and how some of these effects are inhibited or damped depending on the location in the parameter space.
Nevertheless, being able to distinguish between these modes requires to compute the wave action of the system \citep{LK20} or the nonlinear stage of the dynamics. 
Therefore, the next step is to understand the consequences of these new instabilities on the flow dynamics and to determine how they saturate.

A further step would be to perform a weakly nonlinear analysis of the system while taking into account the wave-mean flow interactions (as in \cite{SS71} for instance).
This necessitates the implementation of a multiple scaling approach, alongside the computation of a Ginzburg–Landau equation with cubic or quintic nonlinearity.
In this context, the emergence of a modulational instability is to be expected, as predicted in the classical Benjamin--Feir problem \citep{B67,SD78}.
It may also be of interest to consider the shallow water counterpart and to derive a modified Korteweg--de Vries equation incorporating viscous effects \citep{M81}.
The analytical results, though challenging to obtain, may reveal further insights into the nature of the wave interactions responsible for the radiation-induced instability originally predicted by \cite{N85} in the inviscid limit.

A long-term objective is to compute the nonlinear set of equations by means of direct numerical simulations.
However, solving the Navier--Stokes equations in a domain of unknown shape and with nonlinear boundary conditions has proven to be an exceptionally challenging task for decades \citep{Tuckerman}.
Furthermore, the fluid-structure interaction problem is complicated by the inclusion of terms from differential geometry (see Appendix \ref{app0}), due to the material properties of the membrane.
It is therefore a scientific challenge to provide an efficient method for solving this nonlinear problem in three-dimensional (or even two-dimensional) settings.
\\

\noindent{\bf Acknowledgments.} The author is grateful to D. Clamond for fruitful discussions within the course of this study. \\


\noindent{\bf Declaration of interests.} The author reports no conflict of interest.


\appendix

\section{The velocity and acceleration of moving surfaces}
\label{app0}

Let a non-overturning interface being described by the relation $f(x,z,t) = z + 1 - \zeta(x,t) = 0$ with the associated normal and tangent unit vectors given by \rf{nbtb}.
We assume the surface as having no tangential velocity but a normal velocity defined as
\begin{equation}
V_m^{\perp} = \left. \frac{\partial_t f}{\vert\vert\bm{\nabla}f\vert\vert} \right|_{z=-1+\zeta} = - \frac{\partial_t\zeta}{\sqrt{1+(\partial_x\zeta)^2}} .
\end{equation}

To estimate the Lagrangian derivative $\ud/\ud t$ following the interface motion, we introduce the horizontal and vertical displacements of the interface as, respectively,
\begin{equation}
\frac{\ud x}{\ud t} = V_m^{\perp} ( \bm{n}_{\textrm{b}} \bm{\cdot} \bm{e}_x ) = - \frac{\partial_x\zeta \partial_t\zeta}{1+(\partial_x\zeta)^2} , 
\quad 
\frac{\ud \zeta}{\ud t} = V_m^{\perp} ( \bm{n}_{\textrm{b}} \bm{\cdot} \bm{e}_z ) = \frac{\partial_{t}\zeta}{1+(\partial_x\zeta)^2} ,
\end{equation}
where the expressions are evaluated at the coordinate $z=-1+\zeta(x,t)$.
Alternatively, we can use instead the angle made by the deflection of the surface, i.e. $\tan\theta=\partial_x\zeta$ \citep{Aris}.

The interface acceleration components are thus given by
\begin{align}
A_m^{\perp} &= \bm{n}_{\textrm{b}} \bm{\cdot} \frac{\ud (V_m^{\perp}\bm{n}_{\textrm{b}})}{\ud t} = \frac{\ud V_m^{\perp}}{\ud t} + V_m^{\perp} \left( \bm{n}_{\textrm{b}} \bm{\cdot} \frac{\ud \bm{n}_{\textrm{b}}}{\ud t} \right) = \frac{\ud V_m^{\perp}}{\ud t} , \label{Amn} \\
A_m^{\parallel} &= \bm{t}_{\textrm{b}} \bm{\cdot} \frac{\ud (V_m^{\perp}\bm{n}_{\textrm{b}})}{\ud t} = V_m^{\perp} \left( \bm{t}_{\textrm{b}} \bm{\cdot} \frac{\ud \bm{n}_{\textrm{b}}}{\ud t} \right) = V_m^{\perp} \frac{\ud \theta}{\ud t} , \label{Amt}
\end{align}
since we have
\begin{equation}
\frac{\ud \bm{n}_{\textrm{b}}}{\ud t} = - \bm{t}_{\textrm{b}} \frac{\ud \theta}{\ud t} = \frac{\bm{t}_{\textrm{b}}}{1+(\partial_x\zeta)^2} \left[ \partial_{xt}\zeta - \frac{\partial_x\zeta\partial_t\zeta\partial_{xx}\zeta}{1+(\partial_x\zeta)^2} \right] .
\end{equation}

\section{The inviscid limit and the Nemtsov problem}
\label{appA}

In the diffusionless regime, we set $\Rey\to\infty$ and discard the tangential boundary conditions from \rf{leom1}--\rf{w1w2} as friction is no longer considered.
As in the viscous case, there are infinitely many base flows satisfying the boundary conditions at the surfaces, although we now assume slip condition at the membrane.
We opt for the simplest velocity profile, i.e. the uniform flow $U=\Fr\bm{e}_x$, since it is the configuration originally studied by \cite{N85}.

Eliminating the displacement fields at both interfaces by means of the kinematic conditions, the resulting set of equations reduces to
\begin{alignat}{2}
\label{leominv1}
( \omega - k \Fr ) (\partial_{zz} - k^2) \widehat{w}_1 &= 0 \quad &&\textrm{in} \:\: \Omega_1 , \\
\label{leominv2}
\omega (\partial_{zz} - k^2) \widehat{w}_2 &= 0 \quad &&\textrm{in}  \:\: \Omega_2 ,
\end{alignat}
supplemented with the boundary conditions
\begin{alignat}{2}
(\omega-k\Fr)^2 \partial_{z}\widehat{w}_1 - k^2 (1 + k^2\Bos) \widehat{w}_1 &= 0 \quad &&\textrm{at} \:\: z=0 ,\label{bcinv1}  \\
(\omega-k\Fr)^2 (\alpha_f \partial_{z}\widehat{w}_1 - \partial_{z}\widehat{w}_2) - k^2 (\alpha_m \omega^2 - k^2\Bob + \alpha_f - 1) \widehat{w}_1 &= 0 \quad &&\textrm{at}  \:\: z=-1 , \label{bcinv2}
\end{alignat}
and the continuity of velocity at the lower interface, i.e. $\widehat{w}_1=\widehat{w}_2$ (there is also the condition of evanescence $\widehat{w}_2\to 0$ at $z\to-\infty$).

In the upper layer, the general solution of \rf{leominv1} is given by
\begin{equation}
\label{w1inv}
\widehat{w}_1(z) = A \ue^{kz}  + B \ue^{-kz} ,
\end{equation}
where $A$ and $B$ are arbitrary constants. 
In a similar way, we solve explicitly the problem in the lower fluid layer by integrating equation \rf{leominv2} and applying the appropriate boundary conditions.
It yields
\begin{equation}
\label{w2inv}
\widehat{w}_2(z) = \left.\widehat{w}_1\right|_{z=-1} \ue^{k(z+1)} = A \ue^{kz}  + B \ue^{k(z+2)} .
\end{equation}

Substituting the ansatze \rf{w1inv}--\rf{w2inv} within \rf{bcinv1}--\rf{bcinv2} yields a finite-dimensional quadratic eigenvalue problem of the form \rf{qep} for $\omega$ and $\widehat{\bm{\xi}}=(A,B)^T$.
The associated matrices are
\begin{equation*}
\pazocal{M} = 
\begin{pmatrix}
1 & -1 \\
(\alpha_f-1-k\alpha_m) & -(\alpha_f+1+k\alpha_m) \ue^{2k} 
\end{pmatrix} , 
\quad
\pazocal{S} = -2k\Fr
\begin{pmatrix}
1 & -1 \\
(\alpha_f-1) & -(\alpha_f+1) \ue^{2k} 
\end{pmatrix} ,
\end{equation*}
and
\begin{equation*}
\pazocal{K} = 
\begin{pmatrix}
k(k\Fr^2 - k^2\Bos - 1) & -k(k\Fr^2 + k^2\Bos + 1) \\
k[k^2\Bob + (\alpha_f-1)(k\Fr^2-1)] & k[k^2\Bob - k\Fr^2(\alpha_f+1) - (\alpha_f-1)] \ue^{2k} 
\end{pmatrix} .
\end{equation*}

The dispersion relation of the inviscid two-layer system is obtained by requiring the determinant of the linear system to vanish.
After some algebraic manipulations, it reduces to
\begin{align}
\pazocal{D}_{\textrm{inv}}(\omega,k) = &\left[ (\omega-k\Fr)^2 + k(\alpha_m \omega^2 - k^2\Bob - 1) \right] \left[ (\omega-k\Fr)^2 - k\tanh{(k)} \right] \nonumber \\
&+ \alpha_f \left[ (\omega-k\Fr)^4 - k^2(1+k^2\Bos) \right] \tanh{(k)} - k^3 \alpha_f \Bos (\omega-k\Fr)^2  = 0 . \label{drinv}
\end{align}

In the Nemtsov problem, the flow below the membrane was assumed inviscid, motionless, unperturbed and with the same pressure (or density) as the fluid lying above.
We fix therefore $\widehat{w}_2=0$, $\alpha_f=1$ and $\Bos=0$ (we further assume no surface tension) in the system \rf{bcinv1}--\rf{bcinv2}, while using the same ansatz \rf{w1inv} to compute the associated dispersion relation in the same way as above.
We find
\begin{align}
\pazocal{D}_{\textrm{N}}(\omega,k) = k (\alpha_m \omega^2 - k^2\Bob - 1)  \left[ (\omega-k\Fr)^2 - k\tanh{(k)} \right] + \left[ (\omega-k\Fr)^4 - k^2 \right] \tanh{(k)} = 0 , \nonumber
\end{align}
which reduces to the relation (3.20) from \cite{LK20} when substituting the notations $\alpha_m=1/\alpha$, $\Bob=M_w^2/\alpha$ and $\Fr=M$.

It becomes clear now that our two-layer configuration is a general extension of the paradigmatic model of Nemtsov, taking into consideration the viscous stresses, the influence of shear motion and the interplay with the lower fluid layer.
By relating these two problems, we inevitably gain some physical intuitions on the instability mechanisms into play.

\section{Stability of Rayleigh--Taylor problem with a material interface} 
\label{appB}

We seek to demonstrate the unconditional stability of a material interface enclosed by two semi-infinite viscous fluid layers from the characteristic equation \rf{drm} and under the condition $\alpha_f<1$.
The challenge is to locate every complex roots of this expression and to ensure they all lie in the left half-plane.

First, we notice that the dispersion relation can be rewritten in the form
\begin{equation}
\label{drm2}
\pazocal{D}_m(\Omega) = \left( 1 + \sqrt{\Omega+1} \right) \left[ Q_1(\Omega) + Q_2(\Omega) \right] + \left[ P_1(\Omega) + P_2(\Omega) \right] = 0 ,
\end{equation}
where $Q_1=a_0\Omega^2+a_1\Omega+\epsilon$, $Q_2=a_2-\epsilon$, $P_1=b_0\Omega^2 - \delta$, $P_2=-b_1\Omega-(b_2-\delta)$ and with the introduction of arbitrary small coefficients $0<\delta<\epsilon\ll\vert a_j\vert$.
When $\alpha_f<1$, using term-by-term identification from \rf{drm}, we deduce that the $a_j$ and $b_j$ coefficients are real and positive whatever the values of the other parameters are.
The latter implies that $Q_1$ is a Hurwitz polynomial or equivalently, that its zeros have negative real parts.

Now, let consider a Jordan curve $\mathcal{C}$ consisting in the semi-circle with radius $R$ oriented in the right half-plane.
Allowing the contour to enclose the whole half-plane (by taking the limit $R\to\infty$) it is trivial to show that $\vert Q_1(R\ue^{\ui\phi}) \vert$ (with $-\pi/2\le\phi\le\pi/2$) grows faster than every other terms in \rf{drm2}, since $a_0>b_0$.
It is also direct to prove that $\vert Q_1(\ui x) \vert>\vert P_1(\ui x) \vert$ on the imaginary axis restricted to $x>0$ (due to symmetry).
It follows immediately that $\vert Q_1(z) \vert>\vert P_1(z) \vert$ for $z\in\mathcal{C}$ and thus, by means of Rouch\'e's theorem, that all zeros of $Q_1+P_1$ lie outside the region bounded by the contour $\mathcal{C}$ (since $Q_1$ is Hurwitz stable).
Because the function $h(z) = 1 + \sqrt{z+1}$ is strictly convex and positive on the Jordan curve, the same argument holds for the composition $h(z)Q_1(z)+P_1(z)$ with $z\in\mathcal{C}$.

The next step is to prove the stability of our expression by including the contribution of $h(\Omega)Q_2(\Omega)$.
Doing so is possible from successive applications of Rouch\'e's theorem on the latter expression when being subdivided into a sum of arbitrary small pieces bounded by the linear component of $Q_1$ on $\mathcal{C}$.
Finally, one last application of this theorem on $P_2$ yields, as expected, the proof for unconditional stability of \rf{drm2} when $\alpha_f<1$.

\section{The coefficients of the dispersion relation for the configuration without background flow} 
\label{appC}

We give the explicit expressions of the coefficients $\gamma_j$ from \rf{drfull} as follows
\begin{alignat*}{2}
\gamma_1 &= &&(X^2-1) \big\{ (\alpha_f-1)\left[ 2\alpha_f - (\alpha_f-1) \pazocal{D}_0(X,k) \right] - (\alpha_f+1)(k\alpha_m\mu_1 + k^2\Bob) \big\} \\
& &&+ k\alpha_m(\alpha_f-1)(X-1)^2- 4\mu_1\alpha_f X(X+1) , \\
\gamma_2 &= &&-(X+1) \big\{ (\alpha_f+1)\left[k\alpha_m(X+\mu_1) + k^2\Bob\right] - (\alpha_f-1) \\
& &&\times\left[(X+1)\pazocal{D}_0(X,k) - \alpha_f(X-1)\pazocal{D}_0(-X,-k) + k\alpha_m(X\mu_1-1) + k^2\Bob X\right] \big\}, \\
\gamma_3 &= &&(\alpha_f-1) \big\{ (X^2-1)\left[ (\alpha_f-1)\pazocal{D}_0(X,k) - k\alpha_m\mu_1 - k^2\Bob \right] - k\alpha_m(X-1)^2 \big\} , \\
\gamma_4 &= &&(X+1) \big\{ (\alpha_f^2-1)(X-1)\left[ \pazocal{D}_0(-X,-k) - 2 \right] + (\alpha_f+1) X \left[ k\alpha_m (\mu_1+1) + k^2\Bob \right] \\
& &&- (\alpha_f-1) \left[ k\alpha_m (\mu_1-1) + k^2\Bob + 2(\mu_1+1) \right] \big\} , \\
\gamma_5 &= && 16\mu_2 X (X+1) \big\{ k^2  \Bob -  (\alpha_f-1) \left[ \pazocal{D}_0(X,k) + 2\alpha_f (X-1) \mu_2 \right] \big\} \\
& &&+ 8 k \alpha_m X \mu_2 \left[ \alpha_f (X^2+3) + 2(X+1)\mu_1 + 2(X-1) \right],
\end{alignat*}
where $\pazocal{D}_0(X,k)$ and $\mu_j(X,k)$ are obtained from \rf{dr0} and \rf{mu}, respectively.


\bibliographystyle{jfm}
\bibliography{Biblio}

\end{document}